\pdfoutput=1

\documentclass[12pt,a4paper]{article}

\usepackage{ifthen} 
\newboolean{pdflatex}
\setboolean{pdflatex}{true} 

\newboolean{articletitles}
\setboolean{articletitles}{true} 

\newboolean{uprightparticles}
\setboolean{uprightparticles}{false} 

\def\paperauthors{LHCb collaboration} 

\def\paperasciititle{Searches for rare $B_s^0$ and $B^0$ decays into four muons} 
\def\papertitle{Searches for rare \Bs and \Bd decays into four muons} 
\def\paperkeywords{{High Energy Physics}, {LHCb}} 
\def\papercopyright{\the\year\ CERN for the benefit of the LHCb collaboration} 
\def\paperlicence{CC BY 4.0 licence}
\def\paperlicenceurl{https://creativecommons.org/licenses/by/4.0/}

\usepackage[top=1in, bottom=1.25in, left=1in, right=1in]{geometry}

\usepackage{microtype}
\usepackage{lineno}  
\usepackage{xspace} 
\usepackage{caption} 

\usepackage{graphicx}  
\usepackage{color}
\usepackage{colortbl}
\graphicspath{{./figs/}} 
\usepackage{placeins}

\usepackage{amsmath} 
\usepackage{amssymb}
\usepackage{amsfonts}
\usepackage{upgreek} 

\newcommand*\patchAmsMathEnvironmentForLineno[1]{%
\expandafter\let\csname old#1\expandafter\endcsname\csname #1\endcsname
\expandafter\let\csname oldend#1\expandafter\endcsname\csname
end#1\endcsname
 \renewenvironment{#1}%
   {\linenomath\csname old#1\endcsname}%
   {\csname oldend#1\endcsname\endlinenomath}%
}
\newcommand*\patchBothAmsMathEnvironmentsForLineno[1]{%
  \patchAmsMathEnvironmentForLineno{#1}%
  \patchAmsMathEnvironmentForLineno{#1*}%
}
\AtBeginDocument{%
\patchBothAmsMathEnvironmentsForLineno{equation}%
\patchBothAmsMathEnvironmentsForLineno{align}%
\patchBothAmsMathEnvironmentsForLineno{flalign}%
\patchBothAmsMathEnvironmentsForLineno{alignat}%
\patchBothAmsMathEnvironmentsForLineno{gather}%
\patchBothAmsMathEnvironmentsForLineno{multline}%
\patchBothAmsMathEnvironmentsForLineno{eqnarray}%
}

\usepackage{hyperxmp}

\usepackage[pdftex,
            pdfauthor={\paperauthors},
            pdftitle={\paperasciititle},
            pdfkeywords={\paperkeywords},
            pdfcopyright={Copyright (C) \papercopyright},
            pdflicenseurl={\paperlicenceurl}]{hyperref}

\usepackage[colorinlistoftodos,textsize=scriptsize]{todonotes}

\usepackage[bottom,flushmargin,hang,multiple]{footmisc}

\usepackage[all]{hypcap} 

\usepackage{xspace} 
\usepackage{upgreek}

\def\lhcb   {\mbox{LHCb}\xspace}

\def\MagUp {\mbox{\em Mag\kern -0.05em Up}\xspace}

\ifthenelse{\boolean{uprightparticles}}%
{

 \def\Pmu         {\ensuremath{\upmu}\xspace}

 \def\Ppi         {\ensuremath{\uppi}\xspace}

 \def\Ppsi        {\ensuremath{\uppsi}\xspace}

 \def\PDelta      {\ensuremath{\Delta}\xspace}                 
 \def\PXi         {\ensuremath{\Xi}\xspace}                 
 \def\PLambda     {\ensuremath{\Lambda}\xspace}                 
 \def\PSigma      {\ensuremath{\Sigma}\xspace}                 
 \def\POmega      {\ensuremath{\Omega}\xspace}                 
 \def\PUpsilon    {\ensuremath{\Upsilon}\xspace}

 \def\PB      {\ensuremath{\mathrm{B}}\xspace}                 
                  
 \def\PD      {\ensuremath{\mathrm{D}}\xspace}

 \def\PJ      {\ensuremath{\mathrm{J}}\xspace}                 
 \def\PK      {\ensuremath{\mathrm{K}}\xspace}

 \def\Pb      {\ensuremath{\mathrm{b}}\xspace}

 \def\Pe      {\ensuremath{\mathrm{e}}\xspace}

 \def\Pi      {\ensuremath{\mathrm{i}}\xspace}

 \def\Pp      {\ensuremath{\mathrm{p}}\xspace}

 \def\Ps      {\ensuremath{\mathrm{s}}\xspace}

 \def\thebaroffset{0.0em}
}
{

 \def\Pmu         {\ensuremath{\mu}\xspace}

 \def\Ppi         {\ensuremath{\pi}\xspace}

 \def\Ppsi        {\ensuremath{\psi}\xspace}                 
                  
 \mathchardef\PDelta="7101
 \mathchardef\PXi="7104
 \mathchardef\PLambda="7103
 \mathchardef\PSigma="7106
 \mathchardef\POmega="710A
 \mathchardef\PUpsilon="7107
                  
 \def\PB      {\ensuremath{B}\xspace}                 
                  
 \def\PD      {\ensuremath{D}\xspace}

 \def\PJ      {\ensuremath{J}\xspace}                 
 \def\PK      {\ensuremath{K}\xspace}

 \def\Pb      {\ensuremath{b}\xspace}

 \def\Pe      {\ensuremath{e}\xspace}

 \def\Pi      {\ensuremath{i}\xspace}

 \def\Pp      {\ensuremath{p}\xspace}

 \def\Ps      {\ensuremath{s}\xspace}

 \def\thebaroffset{0.18em}
}
\newcommand{\offsetoverline}[2][\thebaroffset]{\kern #1\overline{\kern -#1 #2}}%

\makeatletter
\ifcase \@ptsize \relax
  \newcommand{\miniscule}{\@setfontsize\miniscule{4}{5}}
\or
  \newcommand{\miniscule}{\@setfontsize\miniscule{5}{6}}
\or
  \newcommand{\miniscule}{\@setfontsize\miniscule{5}{6}}
\fi
\makeatother

\DeclareRobustCommand{\optbar}[1]{\shortstack{{\miniscule (\rule[.5ex]{1.25em}{.18mm})}
  \\ [-.7ex] $#1$}}

\def\epem       {{\ensuremath{\Pe^+\Pe^-}}\xspace}

\def\mup        {{\ensuremath{\Pmu^+}}\xspace}
\def\mun        {{\ensuremath{\Pmu^-}}\xspace} 

\def\mumu       {{\ensuremath{\Pmu^+\Pmu^-}}\xspace}

\def\squark    {{\ensuremath{\Ps}}\xspace}

\def\bquark    {{\ensuremath{\Pb}}\xspace}

\def\pion   {{\ensuremath{\Ppi}}\xspace}

\def\pip    {{\ensuremath{\pion^+}}\xspace}
\def\pim    {{\ensuremath{\pion^-}}\xspace}

\def\kaon    {{\ensuremath{\PK}}\xspace}

\def\KorKbar {\kern \thebaroffset\optbar{\kern -\thebaroffset \PK}{}\xspace}

\def\Kp      {{\ensuremath{\kaon^+}}\xspace}
\def\Km      {{\ensuremath{\kaon^-}}\xspace}

\def\Kstarz  {{\ensuremath{\kaon^{*0}}}\xspace}

\def\D       {{\ensuremath{\PD}}\xspace}

\def\DorDbar {\kern \thebaroffset\optbar{\kern -\thebaroffset \PD}\xspace}

\def\Dp      {{\ensuremath{\D^+}}\xspace}
\def\Dm      {{\ensuremath{\D^-}}\xspace}

\def\DpDm    {\ensuremath{\Dp {\kern -0.16em \Dm}}\xspace}

\def\B       {{\ensuremath{\PB}}\xspace}

\def\BorBbar {\kern \thebaroffset\optbar{\kern -\thebaroffset \PB}\xspace}
\def\Bz      {{\ensuremath{\B^0}}\xspace}

\def\Bd      {{\ensuremath{\B^0}}\xspace}

\def\BdorBdbar {\kern \thebaroffset\optbar{\kern -\thebaroffset \Bd}\xspace}

\def\Bs      {{\ensuremath{\B^0_\squark}}\xspace}

\def\BsorBsbar {\kern \thebaroffset\optbar{\kern -\thebaroffset \Bs}\xspace}

\def\Bds     {{\ensuremath{\B_{(\squark)}^0}}\xspace}

\def\jpsi     {{\ensuremath{{\PJ\mskip -3mu/\mskip -2mu\Ppsi}}}\xspace}
\def\psitwos  {{\ensuremath{\Ppsi{(2S)}}}\xspace}

\def\Y#1S{\ensuremath{\PUpsilon{(#1S)}}\xspace}

\def\proton      {{\ensuremath{\Pp}}\xspace}

\def\Lz          {{\ensuremath{\PLambda}}\xspace}

\def\LorLbar     {\kern \thebaroffset\optbar{\kern -\thebaroffset \PLambda}\xspace}

\def\Lb           {{\ensuremath{\Lz^0_\bquark}}\xspace}

\newcommand{\decay}[2]{\ensuremath{#1\!\to #2}\xspace} 

\def\to                 {\ensuremath{\rightarrow}\xspace}

\def\AT#1     {\ensuremath{A_{\mathrm{T}}^{#1}}\xspace}           

\def\C#1      {\ensuremath{\mathcal{C}_{#1}}\xspace}                       
\def\Cp#1     {\ensuremath{\mathcal{C}_{#1}^{'}}\xspace}                    
\def\Ceff#1   {\ensuremath{\mathcal{C}_{#1}^{\mathrm{(eff)}}}\xspace}        
\def\Cpeff#1  {\ensuremath{\mathcal{C}_{#1}^{'\mathrm{(eff)}}}\xspace}       
\def\Ope#1    {\ensuremath{\mathcal{O}_{#1}}\xspace}                       
\def\Opep#1   {\ensuremath{\mathcal{O}_{#1}^{'}}\xspace}                    

\newcommand{\aunit}[1]{\ensuremath{\text{\,#1}}}       

\newcommand{\tev}{\aunit{Te\kern -0.1em V}\xspace}
\newcommand{\gev}{\aunit{Ge\kern -0.1em V}\xspace}
\newcommand{\mev}{\aunit{Me\kern -0.1em V}\xspace}
\newcommand{\kev}{\aunit{ke\kern -0.1em V}\xspace}
\newcommand{\ev}{\aunit{e\kern -0.1em V}\xspace}
\newcommand{\gevgev}{\ensuremath{\gev^2}\xspace} 
\newcommand{\mevc}{\ensuremath{\aunit{Me\kern -0.1em V\!/}c}\xspace}
\newcommand{\gevc}{\ensuremath{\aunit{Ge\kern -0.1em V\!/}c}\xspace}
\newcommand{\mevcc}{\ensuremath{\aunit{Me\kern -0.1em V\!/}c^2}\xspace}
\newcommand{\gevcc}{\ensuremath{\aunit{Ge\kern -0.1em V\!/}c^2}\xspace}

\def\fb   {\ensuremath{\aunit{fb}}\xspace}
\def\invfb   {\ensuremath{\fb^{-1}}\xspace}

\def\fs   {\aunit{fs}}

\newcommand{\chisq}{\ensuremath{\chi^2}\xspace}

\def\gsim{{~\raise.15em\hbox{$>$}\kern-.85em
          \lower.35em\hbox{$\sim$}~}\xspace}
\def\lsim{{~\raise.15em\hbox{$<$}\kern-.85em
          \lower.35em\hbox{$\sim$}~}\xspace}

\def\sPlot{\mbox{\em sPlot}\xspace}

\def\pt         {\ensuremath{p_{\mathrm{T}}}\xspace}

\def\evtgen     {\mbox{\textsc{EvtGen}}\xspace}

\def\geant      {\mbox{\textsc{Geant4}}\xspace}

\def\photos     {\mbox{\textsc{Photos}}\xspace}

\def\pythia     {\mbox{\textsc{Pythia}}\xspace}

\def\tell1  {TELL1\xspace}
\def\ukl1   {UKL1\xspace}

\newcommand{\ie}{\mbox{\itshape i.e.}\xspace}

\def\mmmm       {{\ensuremath{\Pmu^+\Pmu^-\Pmu^+\Pmu^-}}\xspace}
\def\azero {{\ensuremath{a}}\xspace}

\def\BdsTommmm {\decay{\Bds}{\mmmm}}
\def\BscommadTommmm {\decay{\Bs, \Bz}{\mmmm}}
\def\BsTommmm  {\decay{\Bs}{\mmmm}}
\def\BdTommmm  {\decay{\Bz}{\mmmm}}

\def\BdsTommJpsimm {\decay{\Bds}{\jpsi\left(\mumu\right)\mumu}}
\def\BscommdTommJpsimm {\decay{\Bs, \Bz}{\jpsi\left(\mumu\right)\mumu}}
\def\BsTommJpsimm {\decay{\Bs}{\jpsi\left(\mumu\right)\mumu}}
\def\BdTommJpsimm {\decay{\Bz}{\jpsi\left(\mumu\right)\mumu}}

\def\BdsToaammmm {\decay{\Bds}{\azero\left(\mumu\right)\azero\left(\mumu\right)}}
\def\BsToaammmm {\decay{\Bs}{\azero\left(\mumu\right)\azero\left(\mumu\right)}}
\def\BdToaammmm {\decay{\Bz}{\azero\left(\mumu\right)\azero\left(\mumu\right)}}

\def\BsToJpsiphi {\decay{\Bs}{\jpsi \phi}}
\def\BsToJpsiphimmmm {\decay{\Bs}{\jpsi\left(\mumu\right) \phi\left(\mumu\right)}}
\def\BsToJpsiphimmKK {\decay{\Bs}{\jpsi\left(\mumu\right) \phi\left(\Kp\Km\right)}}

\def\BsToJpsipipi {\decay{\Bs}{\jpsi\left(\mumu\right) \pip\pim}}
\def\BdToJpsiKstarKpi {\decay{\Bz}{\jpsi\left(\mumu\right) \Kstarz \left(\Kp\pim\right)}}
\def\BsTommphimm {\decay{\Bs}{\phi\left(\mumu\right) \mumu}}
\def\BsTophiphimmmm {\decay{\Bs}{\phi\left(\mumu\right) \phi\left(\mumu\right)}}
\def\BsToJpsiKK {\decay{\Bs}{\jpsi\left(\mumu\right) \Kp\Km}}
\def\BsToJpsipipi {\decay{\Bs}{\jpsi\left(\mumu\right) \pip \pim}}
\def\BdToKstarmm {\decay{\Bz}{\Kstarz \left(\Kp\pim\right) \mumu}}
\def\BdTorhomm {\decay{\Bz}{\rho \left(\pip\pim\right) \mumu}}
\def\LbToLambdamm {\decay{\Lb}{\PLambda \left(\proton\pim\right) \mumu}}
\def\LbTopKmm {\decay{\Lb}{\PLambda \left(\proton\Km\right) \mumu}}

\def\JpsiTomm {\decay{\jpsi}{\mumu}}
\def\phiTomm {\decay{\phi}{\mumu}}

\def\bTosll{\decay{\bquark}{\squark \ell^+ \ell^-}} 

\usepackage{cite} 
\usepackage{mciteplus}

\usepackage{longtable} 
\begin{document}
\renewcommand{\thefootnote}{\fnsymbol{footnote}}
\setcounter{footnote}{1}

\begin{titlepage}
\pagenumbering{roman}

\vspace*{-1.5cm}
\centerline{\large EUROPEAN ORGANIZATION FOR NUCLEAR RESEARCH (CERN)}
\vspace*{1.5cm}
\noindent
\begin{tabular*}{\linewidth}{lc@{\extracolsep{\fill}}r@{\extracolsep{0pt}}}
\ifthenelse{\boolean{pdflatex}}
{\vspace*{-1.5cm}\mbox{\!\!\!\includegraphics[width=.14\textwidth]{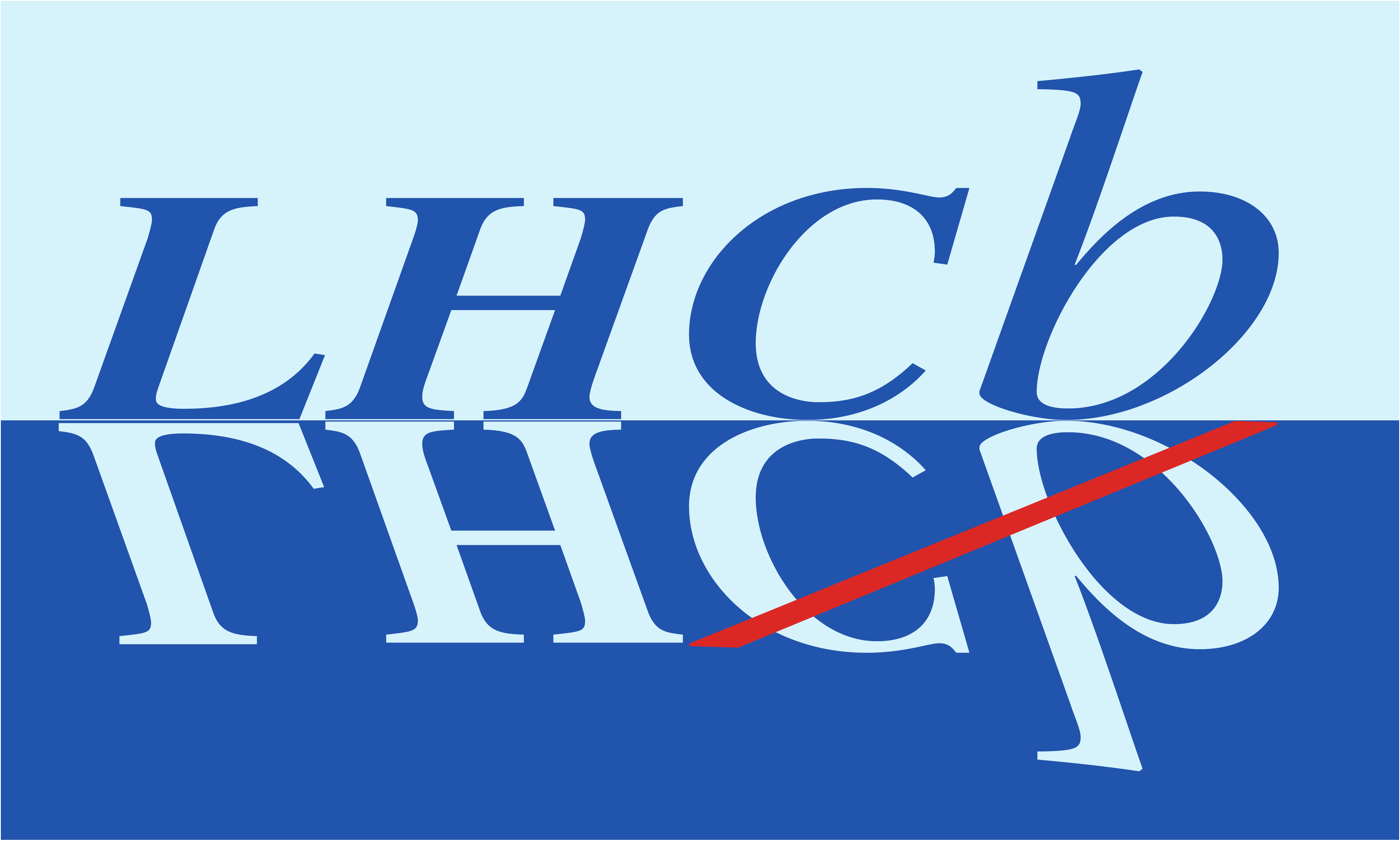}} & &}%
{\vspace*{-1.2cm}\mbox{\!\!\!\includegraphics[width=.12\textwidth]{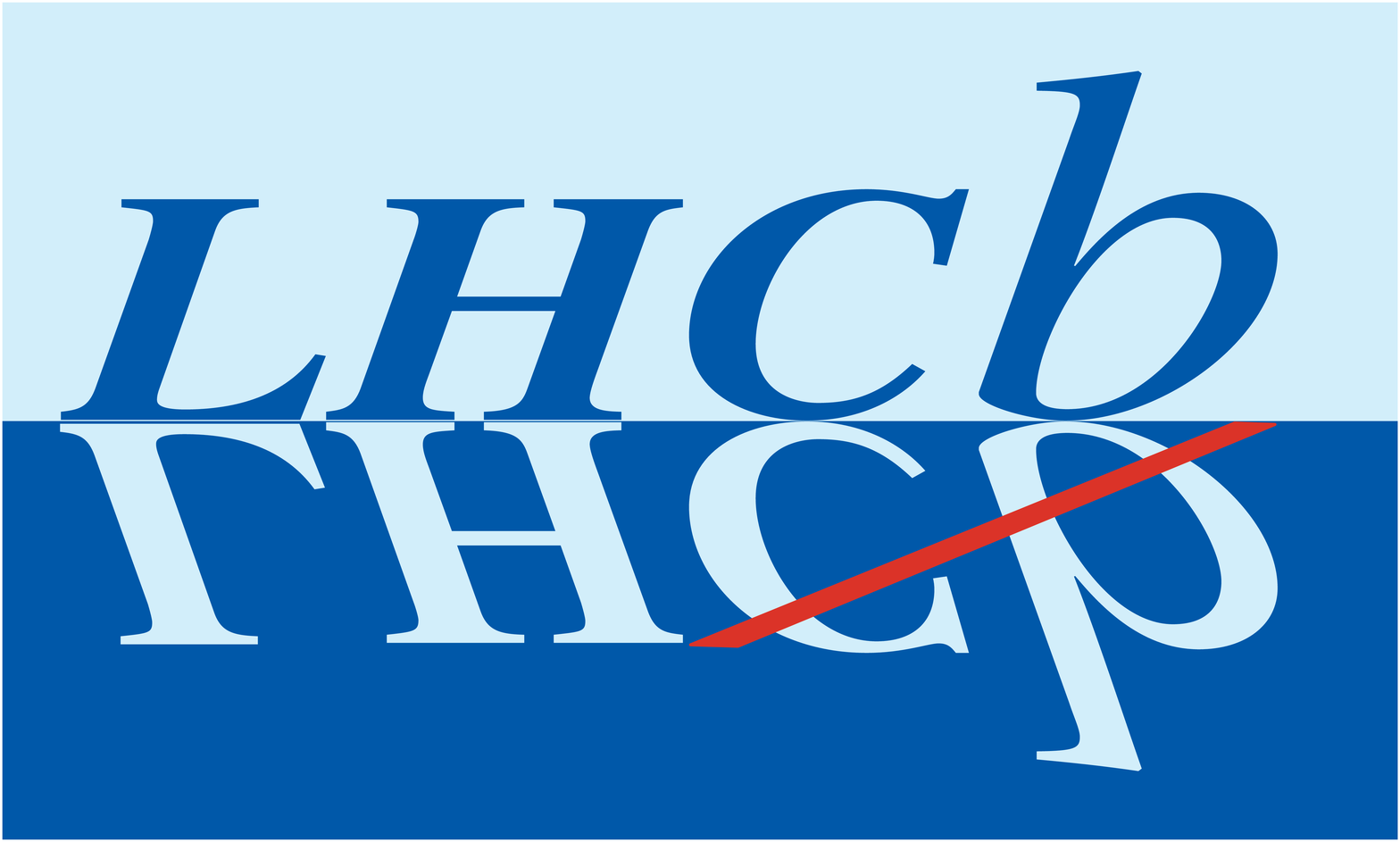}} & &}%
\\
 & & CERN-EP-2021-236 \\  
 & & LHCb-PAPER-2021-039 \\  
 & & March 28, 2022 \\ 
 & & \\
\end{tabular*}

\vspace*{4.0cm}

{\normalfont\bfseries\boldmath\huge
\begin{center}
  \papertitle 
\end{center}
}

\vspace*{2.0cm}

\begin{center}
\paperauthors\footnote{Authors are listed at the end of this paper.}
\end{center}

\vspace*{0.3cm}
This paper is dedicated to the memory of our friend and colleague Alexei Vorobyev.
\vspace*{0.6cm}

\vspace{\fill}

\begin{abstract}
  \noindent
Searches for rare \Bs and \Bd decays into four muons are performed
using proton-proton collision data recorded by the LHCb experiment,
corresponding to an integrated luminosity of 9\invfb . Direct decays and decays via light scalar and \jpsi resonances are considered. No evidence for the six decays searched for is found and upper limits at the 95\% confidence level on their branching fractions ranging between $1.8\times10^{-10}$ and $2.6\times10^{-9}$ are set.

\end{abstract}

\vspace*{2.0cm}

\begin{center}
  Published in JHEP 03 (2022) 109
\end{center}

\vspace{\fill}

{\footnotesize 
\centerline{\copyright~\papercopyright. \href{\paperlicenceurl}{\paperlicence}.}}
\vspace*{2mm}

\end{titlepage}


\newpage
\setcounter{page}{2}
\mbox{~}
%
%
%
%

\renewcommand{\thefootnote}{\arabic{footnote}}
\setcounter{footnote}{0}
\cleardoublepage


\pagestyle{plain} 
\setcounter{page}{1}
\pagenumbering{arabic}

\section{Introduction}
\label{sec:Introduction}
\begin{figure}[b]
\centering
\includegraphics[width=0.75\linewidth]{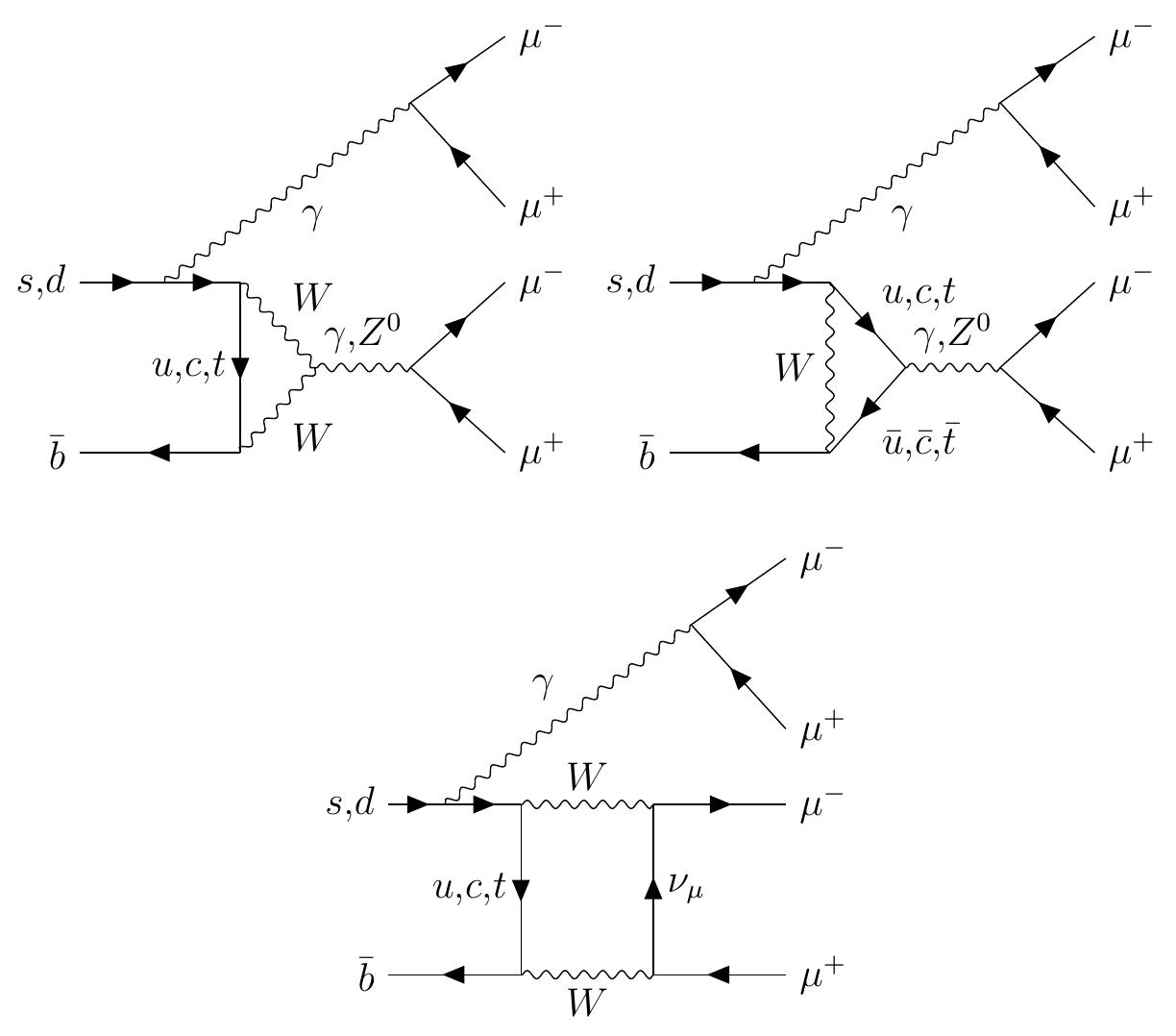}\hfill
\caption{\small Feynman diagrams for nonresonant \BscommadTommmm decays.}\label{fig:feynmandiagrams_b24mu}
\end{figure}

Decays of neutral \Bs and \Bd mesons into four muons that are not mediated by intermediate resonances proceed through $b\rightarrow s$ and $b\rightarrow d$ quark flavour-changing neutral current (FCNC) transitions. These decays proceed by electroweak loop amplitudes in the Standard Model (SM), as illustrated in Fig.~\ref{fig:feynmandiagrams_b24mu}, due to the absence of tree-level FCNCs and hence are highly suppressed, with predicted branching fractions of \mbox{$\mathcal{B}\left( \BsTommmm \right) = \left(0.9-1.0\right) \times 10^{-10}$} and \mbox{$\mathcal{B}\left( \BdTommmm \right) = \left(0.4-4.0\right) \times 10^{-12}$~\cite{Danilina:2018uzr}}.

However, new particles beyond the Standard Model (BSM) may significantly enhance these branching fractions. For example, decays via scalar and pseudoscalar sgoldstino particles into a pair of dimuons in the Minimal Supersymmetric Standard Model (MSSM) may lead to significant enhancements of the branching fractions~\cite{Demidov:2011rd}. Furthermore, rare \Bs and \Bd decays into a pair of dimuons mediated by BSM light narrow scalar resonances ($a$) of the form \BdsToaammmm naturally occur in extensions of the SM involving a new strongly interacting sector. In particular, such models~\cite{Bauer:2017nlg,Liu:2018xkx} can account for the long-standing tension between the SM prediction~\cite{Aoyama:2020ynm} and the observed~\cite{Muong-2:2021ojo,Muong-2:2006rrc} value of the anomalous magnetic dipole moment of the muon, as well as the widely discussed anomalies in \bTosll transitions~\cite{LHCb-PAPER-2021-038,LHCb-PAPER-2021-004,LHCb-PAPER-2019-040,LHCb-PAPER-2017-013,LHCb-PAPER-2021-022,LHCb-PAPER-2020-041,LHCb-PAPER-2020-002}. This motivates a search for $B$ decays into two light scalars with masses around 1\gevcc~\cite{Chala:2019vzu}.
 
This article presents searches for \BdsTommmm decays using proton-proton ($pp$) collision data recorded by the \lhcb experiment in 2011, 2012 (Run 1) and 2015-2018 (Run 2) at centre-of-mass energies of 7, 8 and 13\tev, respectively, corresponding to a total integrated luminosity of 9\invfb. 
Previously, the most sensitive searches for \BdsTommmm\ decays were performed by the \lhcb collaboration using data recorded during Run 1 corresponding to an integrated luminosity of 3\invfb. No signals were observed and the limits $\mathcal{B}\left(\BsTommmm\right) < 2.5\times 10^{-9}$ and $\mathcal{B}\left(\BdTommmm\right) < 6.9\times 10^{-10}$ were set~\cite{LHCb-PAPER-2016-043}. These searches were insensitive to \BdsToaammmm decays with $m_a \approx 1\gevcc$ due to requirements imposed to remove decays via $\phi$ mesons. Now, a specific selection is used that avoids vetoing the dimuon mass region around $1\gevcc$.

\begin{figure}[b]
\centering
\includegraphics[width=0.30\linewidth]{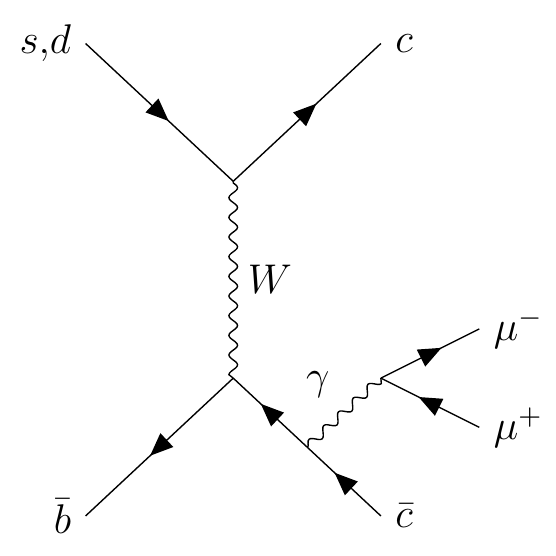}\hfill
\caption{\small Feynman diagram for \BscommdTommJpsimm decays.}\label{fig:feynmandiagrams_b2jpsimm}
\end{figure}

In addition, searches are performed for \BdsTommJpsimm decays, which are tree-level $b\rightarrow c$ transitions that proceed via $W$ boson exchange, as demonstrated in Fig.~\ref{fig:feynmandiagrams_b2jpsimm}.
No SM predictions for these branching fractions are currently available. However, an estimate for the \Bs decay may be obtained using a prediction for the branching fraction of the $\Bs \rightarrow \jpsi \epem$ decay~\cite{Evans:1999zc}, the measured $\jpsi \rightarrow \mup\mun$ branching fraction~\cite{PDG2020}, and assuming that the $\Bs \rightarrow \jpsi \mup \mun$  and $\Bs \rightarrow \jpsi \epem$ branching fractions are equal, yielding $\mathcal{B}\left(\BsTommJpsimm\right) \sim 10^{-11}$. The equivalent \Bz branching fraction may be estimated by multiplying this value by $|V_{cb}V_{cs}|/|V_{cb}V_{cd}| \approx 0.05$, leading to $\mathcal{B}\left(\BdTommJpsimm\right) \sim 10^{-13}$. Since these branching fractions are significantly below the sensitivity of this analysis, the presence of a signal would indicate BSM physics.

The branching fractions are measured using \BsToJpsiphimmmm decays as a normalisation channel, which has a measured branching fraction of \mbox{$\left(1.74\pm 0.14\right) \times 10^{-8}$~\cite{LHCb-PAPER-2020-046, PDG2020}}. Since this decay has the same final state as the signal modes, many sources of systematic uncertainty cancel. First, the yield of the \BsToJpsiphimmmm control mode is estimated using an unbinned maximum-likelihood fit to its invariant mass spectrum. Then, searches for \BdsTommmm\ and \BdsTommJpsimm\ decays are made using unbinned maximum-likelihood fits to the invariant mass spectra of each decay. These are performed simultaneously in multiple intervals of the response of a multivariate classifier constructed to separate signal and combinatorial background, where combinatorial background arises from muons that do not all originate from the same $b$-hadron decay. In contrast, \BdsToaammmm\ decays are searched for by imposing a minimum requirement on the multivariate classifier response due to the very low levels of background, followed by a single fit to the mass spectrum of each signal mode. For all six signal modes, the branching fractions are normalised using the yield of the control mode, its measured branching fraction, and the ratio of efficiencies between the signal and control modes calculated using simulated events. In the case of \Bz decays, the ratio of fragmentation fractions $f_s/f_d$ is also taken into account.

\section{Detector and simulation}
\label{sec:DetectorAndSimulation}

The \lhcb detector is a single-arm forward spectrometer covering the pseudorapidity range \mbox{$2<\eta<5$}, described in detail in Refs.~\cite{LHCb-DP-2008-001,LHCb-DP-2014-002}. It includes a high-precision tracking system consisting of a silicon-strip vertex locator (VELO), surrounding the $pp$ interaction region, a large-area silicon-strip detector located upstream of a dipole magnet with a bending power of about 4\,Tm, and three stations of silicon-strip detectors and straw drift tubes placed downstream of the magnet. Particle identification is provided by two ring-imaging Cherenkov (RICH) detectors, an electromagnetic and a hadronic calorimeter, and a muon system composed of alternating layers of iron and multiwire proportional chambers.

In the simulation, $pp$ collisions are generated using \pythia~\cite{Sjostrand:2007gs,*Sjostrand:2006za} with a specific \lhcb configuration~\cite{LHCb-PROC-2010-056}. Decays of unstable particles are described by \evtgen~\cite{Lange:2001uf}, in which final-state radiation is generated using \photos~\cite{davidson2015photos}. The interaction of the generated particles with the detector, and its response, are implemented using the \geant toolkit~\cite{Allison:2006ve, *Agostinelli:2002hh} as described in Ref.~\cite{LHCb-PROC-2011-006}. Simulated \BdsTommmm decays are generated using a phase-space model for the kinematics of the muons, due to the lack of a SM prediction for the decay dynamics. The \BdsToaammmm decays are simulated with an $a$ state with $m_a = 1\gevcc$, a lifetime of $1\fs$ and a natural width of zero. With these settings it is assumed that the $a$ natural width is significantly below the experimental resolution of approximately 20\mevcc. Simulated \BdsTommJpsimm decays are generated using the \textsc{BTOSLLBALL} model \cite{Ali:1999mm}.

\section{Event selection}
\label{sec:EventSelection}
The online selection is composed of a hardware trigger, followed by a two-stage software trigger. Candidate \BdsTommmm decays are required to pass triggers designed to select decays involving muons, which are identified from tracks that penetrate the calorimeters and the iron layers between the muon stations. In the hardware stage, candidates must pass at least one of two selections: one requiring the presence of at least one high transverse momentum (\pt) muon, the other requiring a pair of muons with a product of their respective transverse momenta above a threshold. In the first software trigger stage, candidates must include a high-\pt muon that is inconsistent with having originated at a primary $pp$ interaction vertex (PV). In the second stage, requirements are imposed on pairs of muons in order to select candidates consistent with $B$ meson decay vertices that are significantly displaced from the PV.

Candidate \BdsTommmm decays are formed by combining four tracks identified as muons that originate from a common decay vertex. The muons forming the candidate are all required to be inconsistent with originating from a PV, have a \pt \textgreater \ 250\mevc, and a good track fit quality. The maximum distance of closest approach between the four tracks is required to be below 0.3 mm. The resulting $B$ candidates are required to have an invariant mass within 1\gevcc of the known \Bs mass~\cite{PDG2020} in the Run 1 dataset or larger than 4\gevcc in the Run 2 dataset, a good quality vertex fit, to be consistent with originating from a PV, have a significant flight distance, and the cosine of the angle between its momentum vector and the vector pointing from the PV to its decay vertex greater than zero. Hadronic background is suppressed using particle identification (PID) information provided by the muon system, calorimeters and RICH detectors, which is used to select well identified muons \cite{LHCb-DP-2013-001}.

The different signal modes and the control channel are separated using requirements on the four invariant masses of pairs of oppositely charged muons $q_{ij}$, where $i$ and $j$ index the positively and negatively charged muons, respectively, such that $ij$ can take the values $\{11, 12, 21, 22\}$. Regions around the $\phi$, \jpsi and \psitwos mesons are defined as \mbox{$950\mevcc < q_{ij} < 1090\mevcc$}, \mbox{$\lvert q_{ij} - m(\jpsi) \rvert < 100\mevcc$} and \mbox{$\lvert q_{ij} - m(\psitwos) \rvert < 100\mevcc$}, respectively, where $m(\jpsi)$ and $m(\psitwos)$ are the known \jpsi and \psitwos masses~\cite{PDG2020}.
The \BdsTommmm signal selection requires that none of the four pairs of opposite-sign muons has an invariant mass within any of the $\phi$, \jpsi or \psitwos regions. This requirement is approximately 64\% efficient on simulated \BsTommmm candidates passing the trigger and offline selection, while rejecting 99.94\% of the \BsToJpsiphimmmm decays. The control mode selection requires that at least one of the mutually exclusive pairs of dimuons (\ie $\mu_1^+\mu_1^-$ and $\mu_2^+\mu_2^-$ or $\mu_1^+\mu_2^-$ and $\mu_2^+\mu_1^-$) has one dimuon mass in the $\phi$ region while the other falls in the \jpsi region. This selection is around 95\% efficient on simulated \BsToJpsiphimmmm decays.

Candidate \BdsToaammmm decays are selected by requiring two mutually exclusive pairs of opposite-sign muons with similar invariant masses, satisfying
\begin{equation}
\label{eq:B2aa_q2}
    \lvert q^2_{ij}-q^2_{kl} \rvert < 2\sqrt{\sigma^2(q^2_{ij}) + \sigma^2(q^2_{kl})} \,,
\end{equation}
where $\sigma(q^2_{ij})$ is the dimuon invariant mass squared resolution evaluated at $q^2_{ij}$. The resolution increases roughly linearly as a function of $q^2$, from around 0.003 \gevgev to around 0.15 \gevgev, and is estimated as a function of $q^2$ separately in each data-taking year using simulated \BsTommmm decays. This requirement does not completely eliminate contamination from \BsToJpsiphimmmm decays, which may satisfy the requirement if muons from the intermediate \jpsi and $\phi$ mesons are paired incorrectly, giving two intermediate combinations with similar masses. This remaining background is vetoed by requiring that none of the four pairs of oppositely charged muons has an invariant mass in the \jpsi region.

The \BdsTommJpsimm candidates are selected by requiring that at least one of the pairs of opposite-sign muons has a mass in the \jpsi region. Meanwhile the corresponding opposite-sign pair of muons is required to have a mass that falls outside the $\phi$ region under both a dimuon and dikaon mass hypothesis, in order to remove background from \BsToJpsiphimmmm and \BsToJpsiphimmKK decays, respectively. To suppress additional hadronic background such as \BsToJpsipipi and \BdToJpsiKstarKpi decays, the two muons not forming the \jpsi candidate are required to satisfy more stringent PID criteria.

Finally, a multivariate classifier based on a boosted decision tree (BDT) algorithm from the TMVA package~\cite{Hocker:2007ht} is used to separate signal from combinatorial background. The classifier is trained using a mixture of simulated \mbox{\BsTommmm} and \mbox{\BdTommmm} decays as the signal proxy, while the background proxy comprises data with an invariant $B$ mass, $m(\mumu\mumu)$, in the regions \mbox{$m(\mumu\mumu) > 5426\mevcc$} and \mbox{$m(\mumu\mumu) < 5020\mevcc$}. The simulated \BdsTommmm decays used in the training are corrected to improve agreement with data as described in Sec.~\ref{sec:Normalisation}. Two separate BDT classifiers are trained for the Run 1 and Run 2 datasets due to the different data-taking conditions. The size of the available training sample is maximised in each case using the k-folding technique \cite{Blum:1999:BHB:307400.307439}, which ensures that the BDT used to classify a given event in the data background sample is trained independent of the event itself. The following properties of the $B$ candidate are taken as inputs to the classifiers: the logarithm of its impact parameter \chisq with respect to the associated PV, the logarithm of its flight distance \chisq, its flight distance, pseudorapidity, \pt, decay time, decay vertex \chisq per degree of freedom, and the minimum impact parameter \chisq with respect to the associated PV of the four muons. Here, the impact parameter \chisq is defined as the difference in the vertex-fit \chisq of a given PV reconstructed with and without the track or candidate being considered. In \mbox{Run 2}, two variables reflecting the isolation of the $B$ candidate from other tracks in the event are also included in the training: the changes in the decay vertex \chisq when either one or two of the closest additional tracks in the underlying event are added to the decay vertex fit. Finally, the classifier response is transformed to give a uniform distribution between 0 and 1 for simulated \BsTommmm candidates.

In the searches for \BdsTommmm and \BdsTommJpsimm decays the data are split into four intervals in the BDT classifier response with boundaries 0.2, 0.3, 0.5, 0.6 and 1.0, and a simultaneous fit to all four BDT intervals is used to search for the decays. Due to the low background levels in the \BdsToaammmm search region, the BDT response in this case is only required to be greater than 0.1. Similarly, the selection for \BsToJpsiphimmmm decays requires the BDT classifier response to exceed 0.05. 

\section{Background}
\label{sec:Background}
A range of background sources have the potential to contaminate the signal search regions. The four-muon decays \BsToJpsiphimmmm and \BsTommphimm, which have branching fractions of $1.74 \times 10^{-8}$~\cite{LHCb-PAPER-2020-046} and $2.3 \times 10^{-10}$~\cite{PDG2020}, respectively, are reduced to negligible levels by the $\phi$ and \jpsi vetoes. Potential background to the \mbox{\BdsToaammmm} search comes from \BsTophiphimmmm decays, however the corresponding branching fraction of $1.5\times 10^{-12}$ is well below the sensitivity of this search.

A second category of background arises from decays of $b$-hadrons of the form \mbox{$H_b \rightarrow h^+ h^{'-} \mup \mun$} where two hadrons ($h^+$ and  $h^{'-}$) are misidentified as muons. Such background includes \mbox{\BsToJpsiphimmKK}, \mbox{\BsToJpsiKK}, \mbox{\BsToJpsipipi}, \mbox{\BdToJpsiKstarKpi}, \mbox{\BdToKstarmm}, \mbox{\BdTorhomm}, \mbox{\LbToLambdamm} and \mbox{\LbTopKmm} decays. The typical probability for a hadron to be misidentified as a muon by the \lhcb detector is around 1\% \cite{LHCb-DP-2013-001}, and this rate is further reduced by the PID requirements described in Sec.~\ref{sec:EventSelection}. Studies performed using simulation demonstrate that all these decays are reduced to negligible levels by the signal and the control sample selections. As such, this analysis is essentially free of background from other $b$-hadron decays, which are therefore not modelled in the fits for the signal and control decays.

\section{Fit procedure}
\label{sec:Normalisation}
The signal yields in the branching fraction measurements are normalised with respect to the \BsToJpsiphimmmm control mode such that the branching fraction of a given signal decay is calculated as
\begin{equation}
    \mathcal{B}_{\mathrm{sig}} = \alpha^{n}_{s/d}\times N^{n}_{\mathrm{sig}},
    \label{eq:norm}
\end{equation}
where the normalisation factors for \Bs and \Bz modes, $\alpha^{n}_{s}$ and $\alpha^{n}_{d}$, are defined as
\begin{align}
    \alpha^{n}_{s} &\equiv \frac{\mathcal{B}_{\text{norm}}}{N_{\text{norm}}}\cdot\frac{\epsilon_{\text{norm}}}{\epsilon_{\text{sig}}^{n}}, \nonumber \\
     \alpha^{n}_{d} &\equiv \frac{\mathcal{B}_{\text{norm}}}{N_{\text{norm}}}\cdot\frac{\epsilon_{\text{norm}}}{\epsilon_{\text{sig}}^{n}}\cdot\frac{f_{s}}{f_{d}}.
\end{align}
The symbols sig and norm refer to the signal and normalisation decays, respectively, $\mathcal{B}$ is the branching fraction, $N$ is the yield, $\epsilon$ is the total efficiency to reconstruct and select a given decay mode and $n$ indexes the BDT response intervals used in the simultaneous fits for \mbox{\BdsTommmm} and \BdsTommJpsimm\ decays (in the case of \mbox{\BdsToaammmm} there is only a single BDT response interval).  

The branching fraction of the \BsToJpsiphimmmm normalisation mode is calculated from the product of the branching fractions of the \BsToJpsiphi~\cite{LHCb-PAPER-2020-046}, \mbox{\JpsiTomm} and \mbox{\phiTomm}~\cite{PDG2020} decays, yielding \mbox{$\mathcal{B}_{\mathrm{norm}} = \left(1.74\pm 0.14\right)\times 10^{-8}$}. The ratio of fragmentation fractions is calculated as a weighted average of values measured at centre-of-mass energies of 7, 8 and 13\tev~\cite{LHCb-PAPER-2020-046}, giving \mbox{$f_s/f_d = 0.250 \pm 0.008$}. The correlation in the measurements of \mbox{$\mathcal{B}\left(\BsToJpsiphi\right)$} and $f_s/f_d$ is taken into account when calculating the uncertainty in the normalisation term $\alpha_{d}^i$. 

The efficiencies are calculated using simulated decays, to which weights are applied in order to improve concordance with data. These weights are calculated by comparing \BsToJpsiphimmKK decays in data and simulation, for which the branching fraction is three orders of magnitude larger than for \mbox{\BsToJpsiphimmmm} and therefore allows for a much more precise determination of differences between data and simulation. The trigger and offline selections for these decays are similar to those used for \BsToJpsiphimmmm decays, but no PID requirement is applied on the kaons. The distributions of the variables of interest for \mbox{\BsToJpsiphimmKK} decays are separated from background in data using the \sPlot method~\cite{Pivk:2004ty}, using the $B$ candidate invariant mass as the discriminating variable. 

A first set of weights, referred to as the generator weights, corrects the \pt and pseudorapidity distributions of $B$ mesons as generated by \pythia~8~\cite{Sjostrand:2007gs}, along with the multiplicity of the underlying events. A second set of weights, referred to as the reconstruction weights, is used to correct the vertex \chisq per degree of freedom and the impact parameter \chisq of the $B$ mesons. The efficiencies for each mode in each data-taking year are then calculated as 
\begin{equation}
    \epsilon = \epsilon_{\text{presel}} \cdot \epsilon_{\text{BDT}} = \frac{\sum\limits_{\text{presel}} \omega_{\text{gen}}}{\sum\limits_{\text{gen}} \omega_{\text{gen}}} \cdot \frac{\sum\limits_{\text{BDT}} \omega_{\text{gen}}\omega_{\text{rec}}}{\sum\limits_{\text{presel}} \omega_{\text{gen}} \omega_{\text{rec}}} \,,
    \label{eq:eff_calculation}
\end{equation}
where  $\sum\limits_{\text{gen}}$, $\sum\limits_{\text{presel}}$ and $\sum\limits_{\text{BDT}}$ refer to sums over the generated simulation sample before any reconstruction or selection, the offline reconstructed simulation sample passing the full selection excluding the BDT requirements, and the offline reconstructed simulation sample passing the full selection including the BDT requirements, respectively. Meanwhile, $\omega_{\text{gen}}$ and $\omega_{\text{rec}}$ refer to the generator and reconstruction weights. Note that the reconstruction weights are only used in the calculation of the efficiencies of the BDT requirements, since requirements on the $B$ candidate's vertex \chisq per degree of freedom and the impact parameter \chisq imposed in prior stages of the selection are highly efficient for the signal modes. The individual efficiencies for each year are then combined in a weighted average according to the integrated luminosity recorded and the $b\overline{b}$ production cross-section in that year~\cite{LHCb-PAPER-2016-031, LHCb-PAPER-2020-018}.

\begin{figure}[b]
    \centering
     \includegraphics[width=0.65\linewidth]{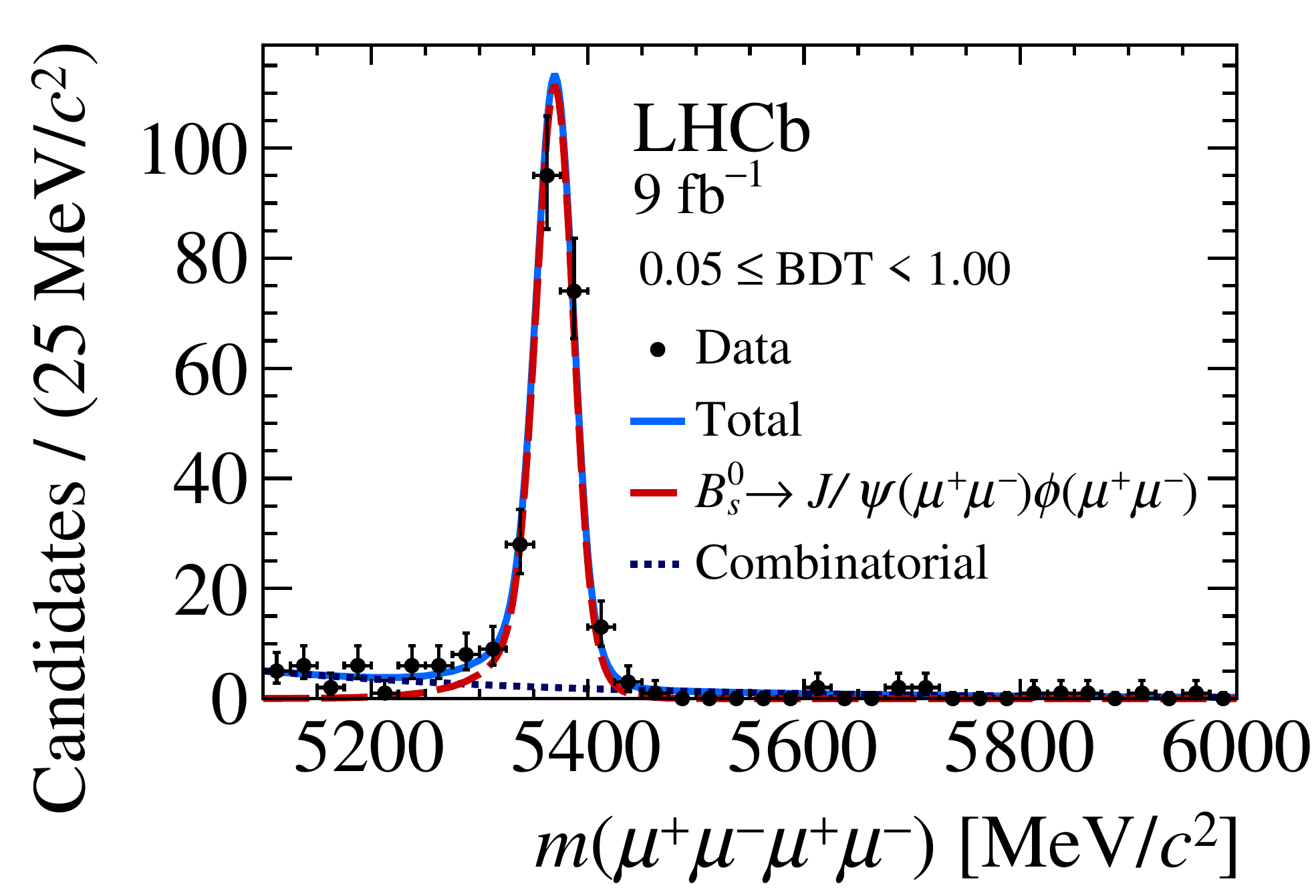}\hfill
\caption{\small Distribution of the $\mumu\mumu$ invariant mass of candidates passing the \mbox{\BsToJpsiphimmmm} selection, with the fit model used to determine the control sample yield overlaid.}
\label{fig:controlMassFit}
\end{figure}

The invariant mass distribution of \BsToJpsiphimmmm candidates in the range \mbox{$5100 < m(\mumu\mumu) < 6000\mevcc$} is shown in Fig. \ref{fig:controlMassFit}. The yield of the normalisation mode is determined using an extended unbinned maximum-likelihood fit to this invariant mass spectrum. The \BsToJpsiphimmmm signal is modelled using the sum of two Crystal Ball functions~\cite{Skwarnicki:1986xj} with common mean, referred to as a double Crystal Ball (DCB) function. The parameters of this function are determined from a fit to simulated \BsToJpsiphimmmm decays, in which the values are fixed in the fit to data, with the exception of the mean and widths of the DCB function, which are allowed to vary freely in order to account for differences in resolution and mass scale between data and simulation. The resulting ratio of the mass resolution in data over simulation is found to be $1.11 \pm 0.09$. The combinatorial background is modelled with an exponential function. Background from the six signal decays in the control mode mass distribution is negligible due to the low branching fractions of the signal decays and the requirements on the \jpsi and $\phi$ masses imposed in the control mode selection. The yield of \mbox{\BsToJpsiphimmmm} decays is found to be $218 \pm 16$, where the uncertainty is statistical only. The invariant mass spectrum and the fit projection are shown in Fig.~\ref{fig:controlMassFit}. 

As described in Sec.~\ref{sec:EventSelection}, the data used to search for \mbox{\BdsTommmm} and \mbox{\BdsTommJpsimm} decays are split into four intervals in the BDT classifier response, while for the \BdsToaammmm data sample a single BDT interval is defined. The normalisation factors $\alpha_s^i$ and $\alpha_d^i$ are calculated using Eq.~\ref{eq:norm} and the inputs described above, with their values Gaussian constrained to their central values according to their statistical uncertainties. Additional Gaussian constraints are used to include the effects of the systematic uncertainties in the inputs described in Sec.~\ref{sec:SystematicUncertainties}.

For each of the \mbox{\BdsTommmm} and \mbox{\BdsTommJpsimm} signal modes, an unbinned extended maximum-likelihood fit is performed to the $B$ candidate invariant mass distribution in the range \mbox{$4900 < m(\mumu\mumu) < 6000\mevcc$}, simultaneously in all four BDT intervals, in order to search for the decays and measure their branching fractions. In the case of the \BdsToaammmm modes, fits are performed to the invariant mass distribution in a single region of the BDT response. All signal mass shapes are modelled using DCB functions, in which the parameters are fixed to values obtained from simulation, with the exception of the mean and widths, which are offset and scaled respectively according to the results of the fit for the control mode. Due to their overlap, the fits to the \Bs and \Bd decays are performed separately in each case, \ie the \Bs component is set to zero when performing the fit to the \Bz component and vice versa. The yields of each signal mode are not fitted directly, rather they are expressed in terms of the normalisation factors and the decay branching fractions (see Eq.~\ref{eq:norm}), which are allowed to vary freely. The branching fraction is constrained to be greater than or equal to zero in each fit. The combinatorial background is modelled using exponential functions, in which the slope parameters and yields are allowed to vary freely independently in each BDT interval. 

\section{Systematic uncertainties}
\label{sec:SystematicUncertainties}
Due to the similarity between the signal and control decay modes, many sources of systematic uncertainty cancel in the ratio of efficiencies. Nonetheless, a number of sources of systematic uncertainty have the potential to affect the measured branching fractions. Two sources of uncertainty arise from the models used to generate the simulated signal decays. In particular, since no theoretical prediction for the decay dynamics of \BdsTommmm decays is currently available, decays are generated using a phase-space model. This may lead to a bias in the efficiency calculation if the true kinematic distributions of these decays differs from the one used in the simulation. To estimate the representative scale of the bias arising from such an effect, the dimuon invariant mass squared ($q^2_{ij}$) distributions for \BdsTommmm decays are weighted to be uniform in the $(q^2_{11},q^2_{22})$ plane, and the efficiencies are evaluated. A similar effect is evaluated for \BdsTommJpsimm decays, where the invariant mass squared of the two muons not originating from the \jpsi meson is weighted to give a uniform distribution. This leads to shifts relative to the nominal efficiencies of up to about 20\% depending on the decay mode and BDT interval, which are taken as systematic uncertainties. This represents the largest source of systematic uncertainty. Similarly, the effective lifetimes of the \Bs meson decays, \mbox{\BsTommmm}, \mbox{\BsToaammmm} and \mbox{\BsTommJpsimm}, depend on the CP-even and CP-odd contributions in the decay amplitudes, which are \textit{a priori} unknown. These decays are generated assuming the mean \Bs lifetime. If their true effective lifetimes differ significantly from this value, then the efficiency estimates will be biased. The maximum size of such an effect is estimated by weighting simulated decay modes to have their effective lifetimes equal to the lifetimes of the heavy and light \Bs mass eigenstates. The largest shift with respect to the nominal efficiency is again taken as a source of systematic uncertainty, with relative shifts in the efficiencies at around the 5\% level.

Further sources of uncertainty arise due to data-simulation differences. The effect of mismodelling of the PID response in simulation is evaluated using calibration samples of muons from data to compare data-driven efficiencies with those obtained directly from simulation, with the differences in the ratios of efficiencies of 1--2\% taken as systematic uncertainties. The effect of the difference in mass resolution between simulation and data on the selection efficiency is estimated to be below 1\% using the results of the fits to the control mode mass distribution. Finally, the efficiencies are evaluated without the application of the generator and reconstruction weights described in Sec.~\ref{sec:Normalisation} in order to get an upper bound on the likely effect of mismodelling in simulation. Since the effect on the efficiency ratio is found to be small, up to around 10\% depending on the decay and BDT interval, this shift is conservatively taken as a source of systematic uncertainty in the efficiency. All these systematic uncertainties are included in the branching fraction fit by applying Gaussian constraints to the efficiencies or efficiency ratios.

\section{Results}
\label{sec:Results}
\begin{figure}[!ht]
    \centering
    \includegraphics[width=0.5\linewidth]{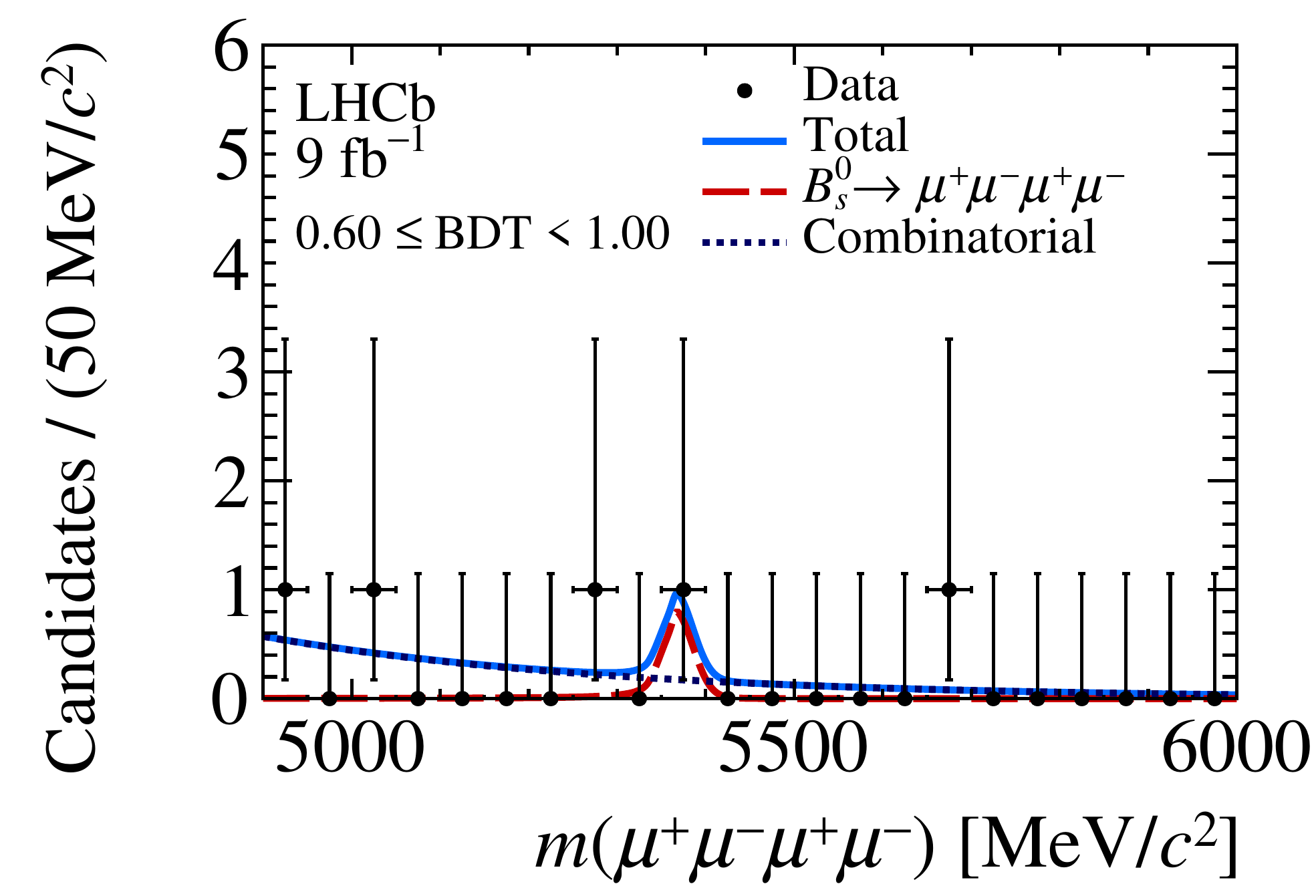}\hfill
     \includegraphics[width=0.5\linewidth]{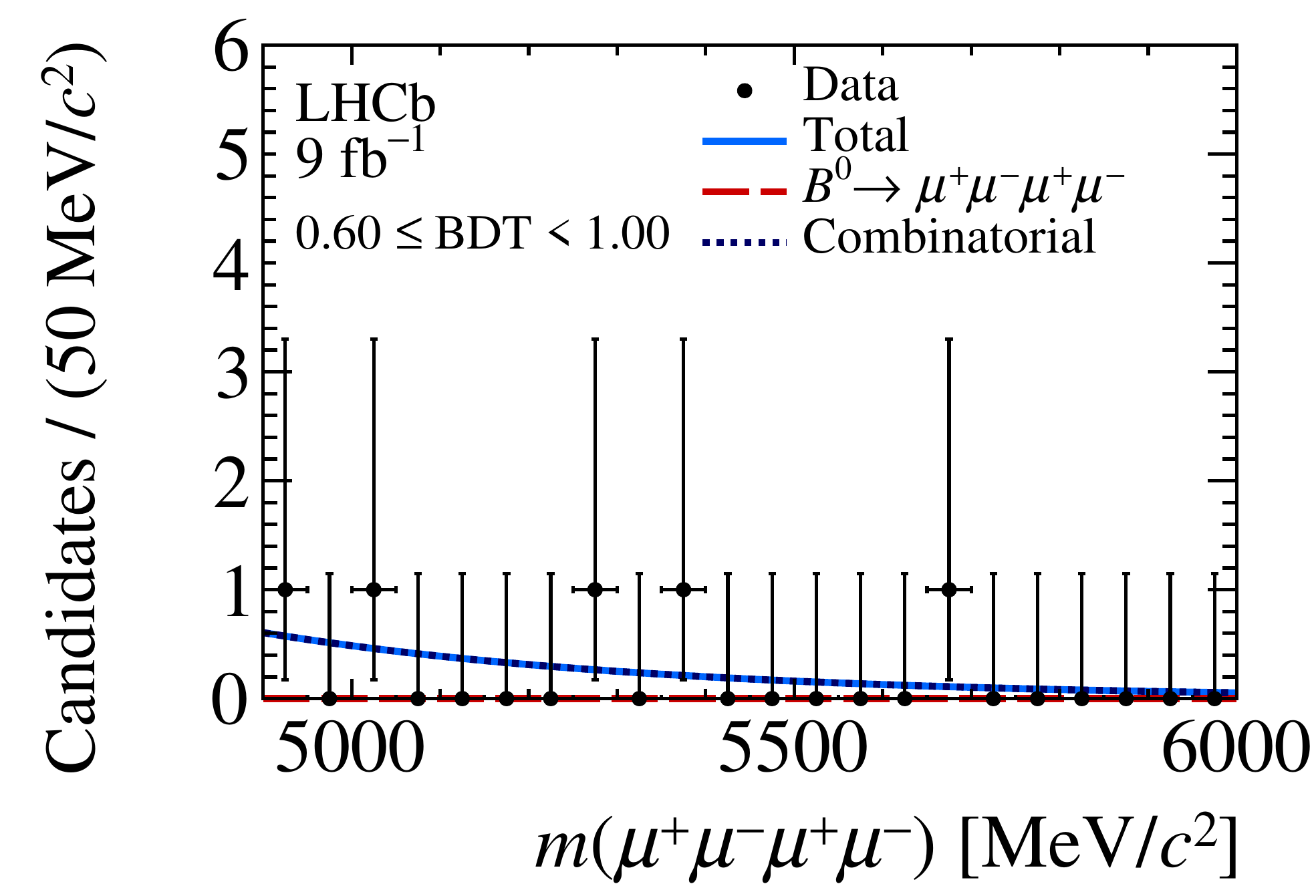}\hfill
\caption{\small Distribution of the $\mumu\mumu$ invariant mass of candidates passing the \mbox{\BdsTommmm} selection in the most sensitive BDT interval, with the fit models used to determine the branching fractions of (left) \mbox{\BsTommmm} and (right) \mbox{\BdTommmm} decays overlaid.}
\label{fig:BdsTommmmMassFit}
\end{figure}

\begin{figure}[!ht]
    \centering
         \includegraphics[width=0.5\linewidth]{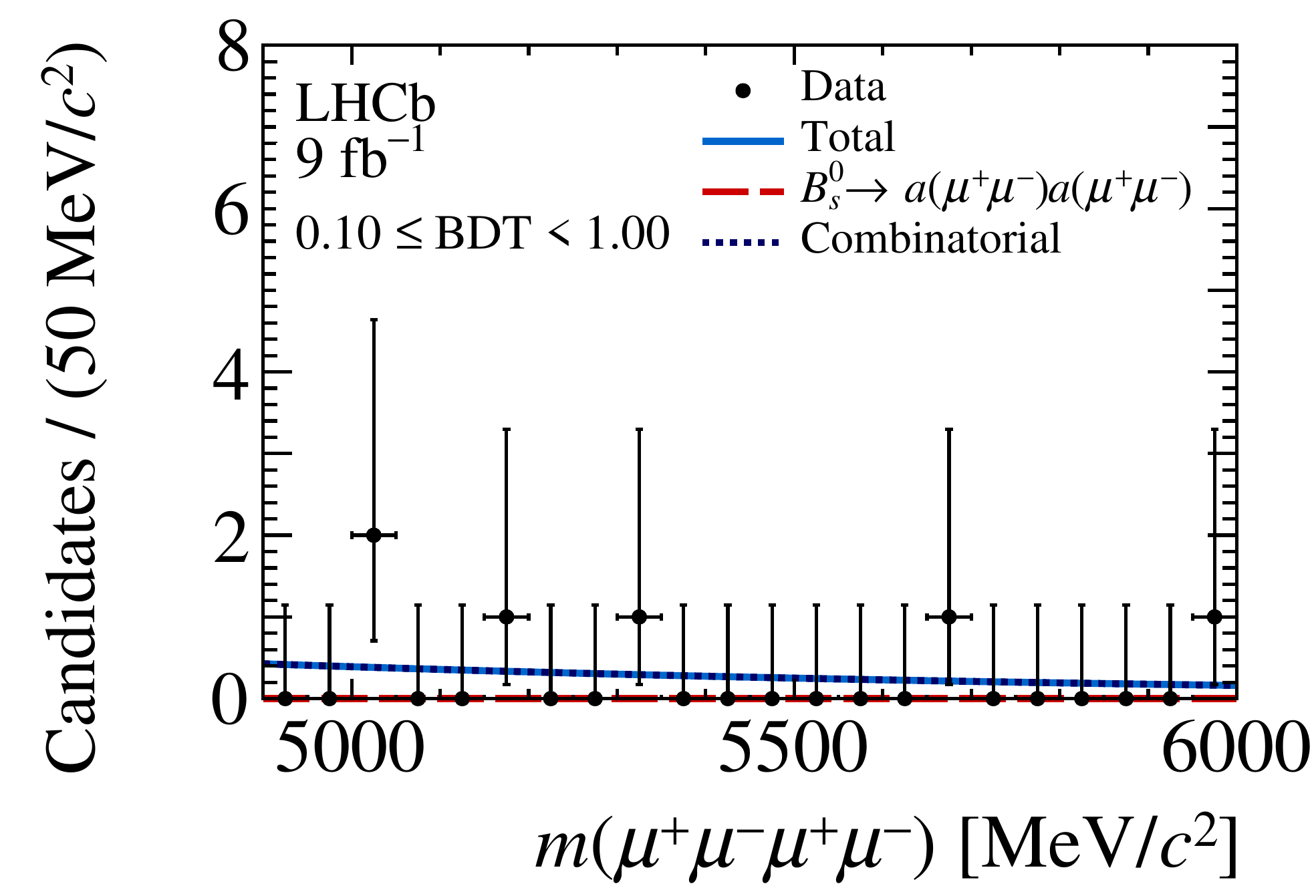}\hfill
     \includegraphics[width=0.5\linewidth]{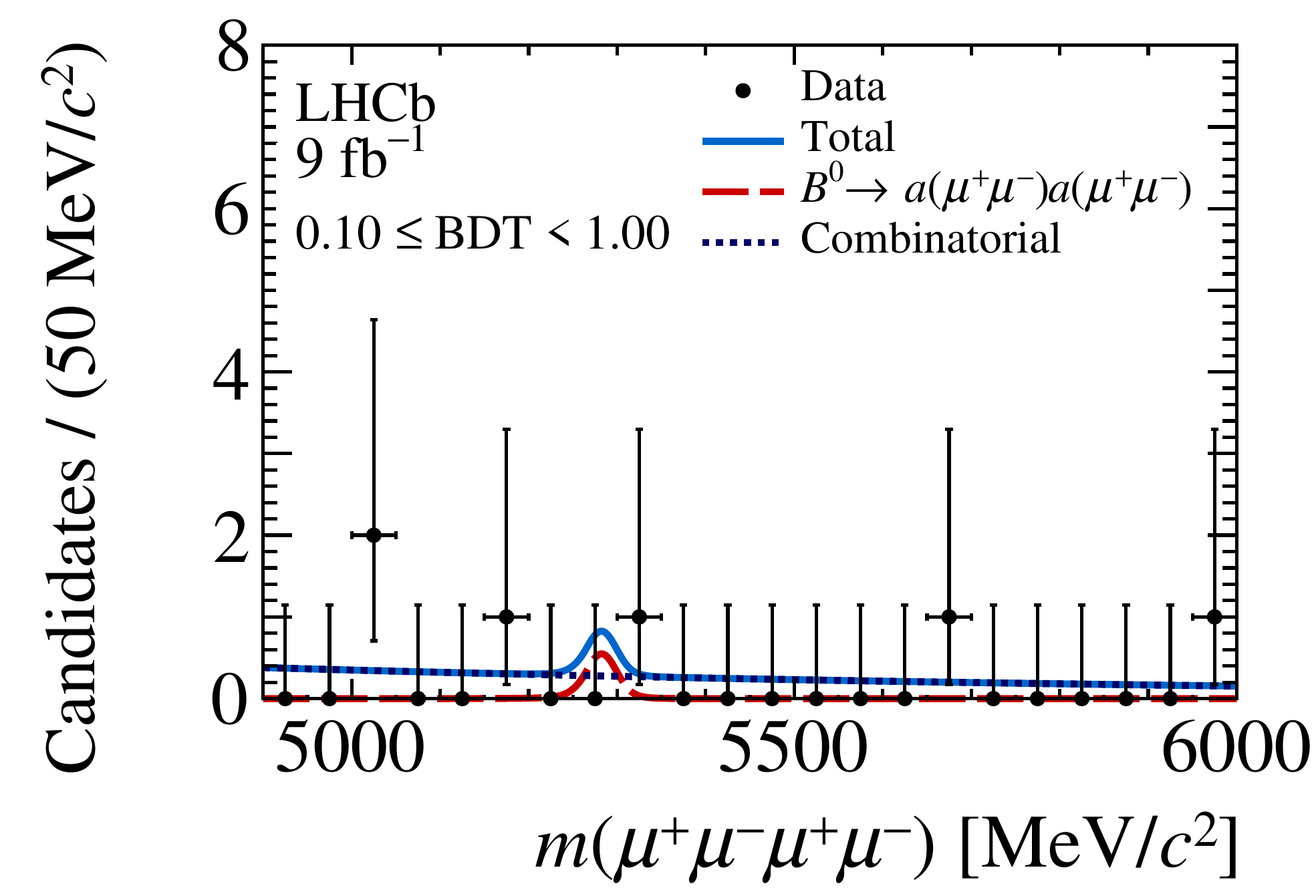}\hfill
 \caption{\small Distribution of the $\mumu\mumu$ invariant mass of candidates passing the \mbox{\BdsToaammmm} selection, with the fit models used to determine the branching fractions of (left) \mbox{\BsToaammmm} and (right) \mbox{\BdToaammmm} decays overlaid.}
\label{fig:BdsToaammmmMassFit}
\end{figure}

\begin{figure}[!ht]
    \centering
    \includegraphics[width=0.5\linewidth]{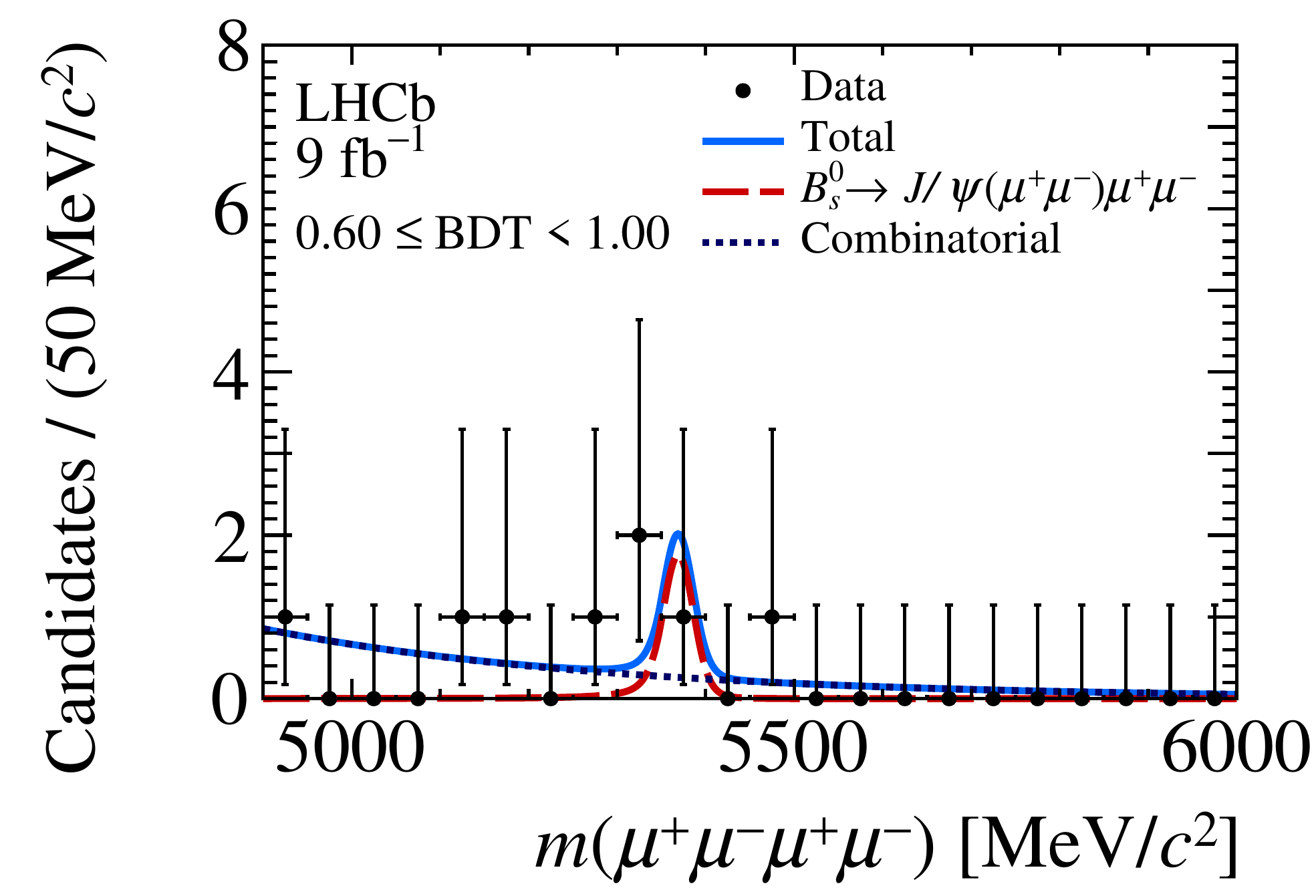}\hfill
     \includegraphics[width=0.5\linewidth]{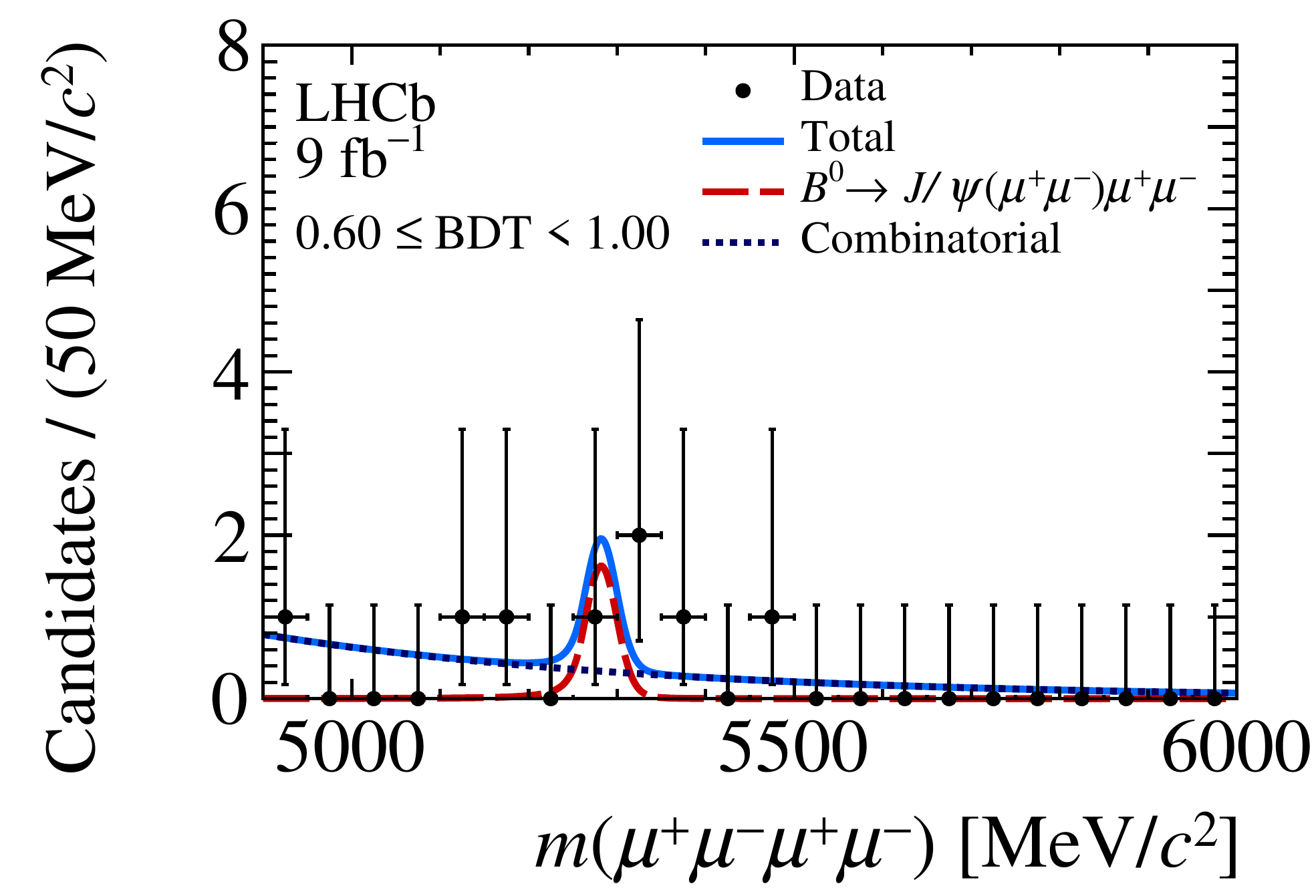}\hfill
\caption{\small Distribution of the $\mumu\mumu$ invariant mass of candidates passing the \mbox{\BdsTommJpsimm} selection in the most sensitive BDT interval, with the fit models used to determine the branching fractions of (left) \mbox{\BsTommJpsimm} and (right) \mbox{\BdTommJpsimm} decays overlaid.}
\label{fig:BdsTommJpsimmMassFit}
\end{figure}

\begin{figure}[!ht]
    \centering
     \includegraphics[width=0.5\linewidth]{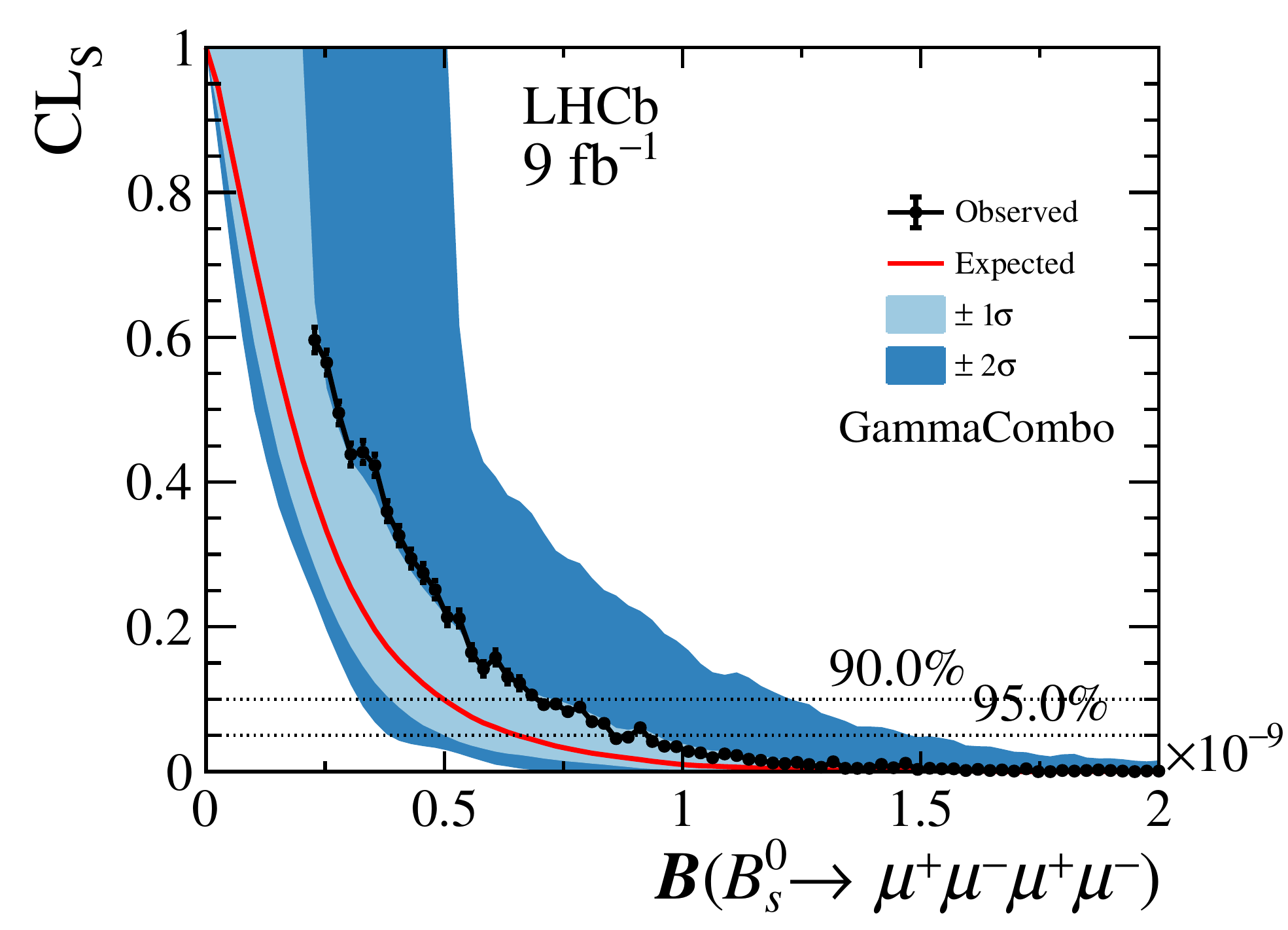}\hfill
     \includegraphics[width=0.5\linewidth]{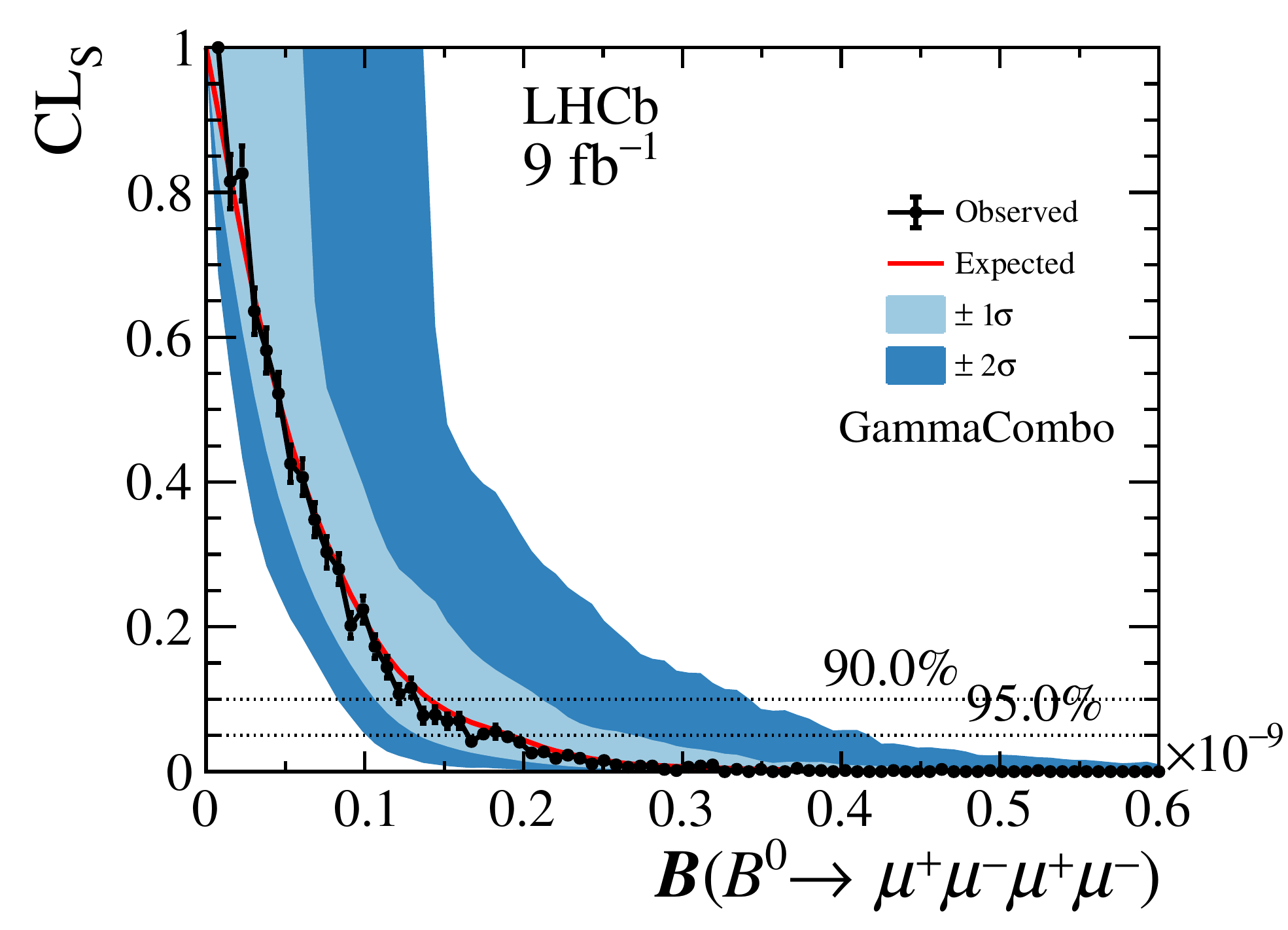}\hfill
     \includegraphics[width=0.5\linewidth]{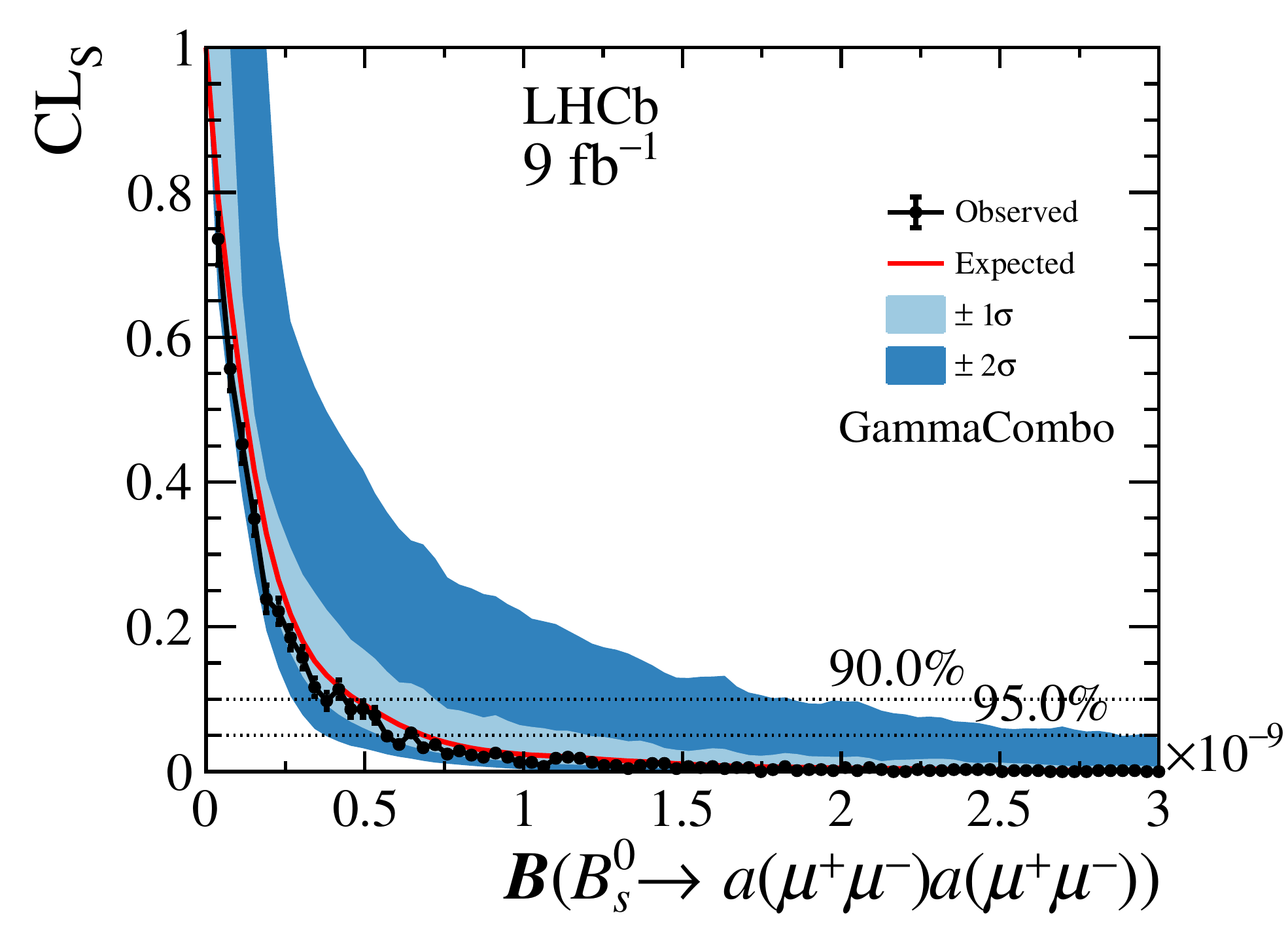}\hfill
     \includegraphics[width=0.5\linewidth]{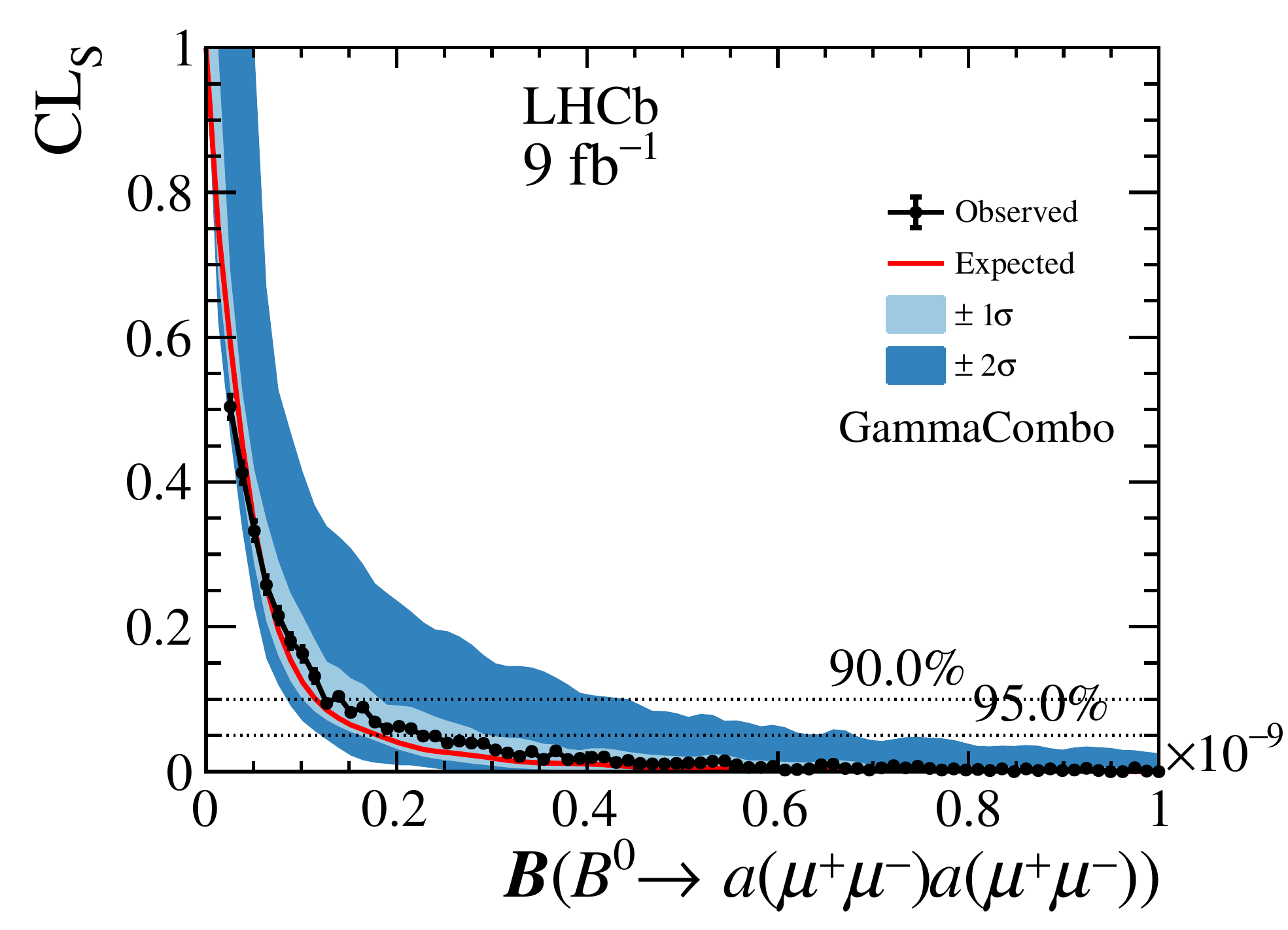}\hfill
     \includegraphics[width=0.5\linewidth]{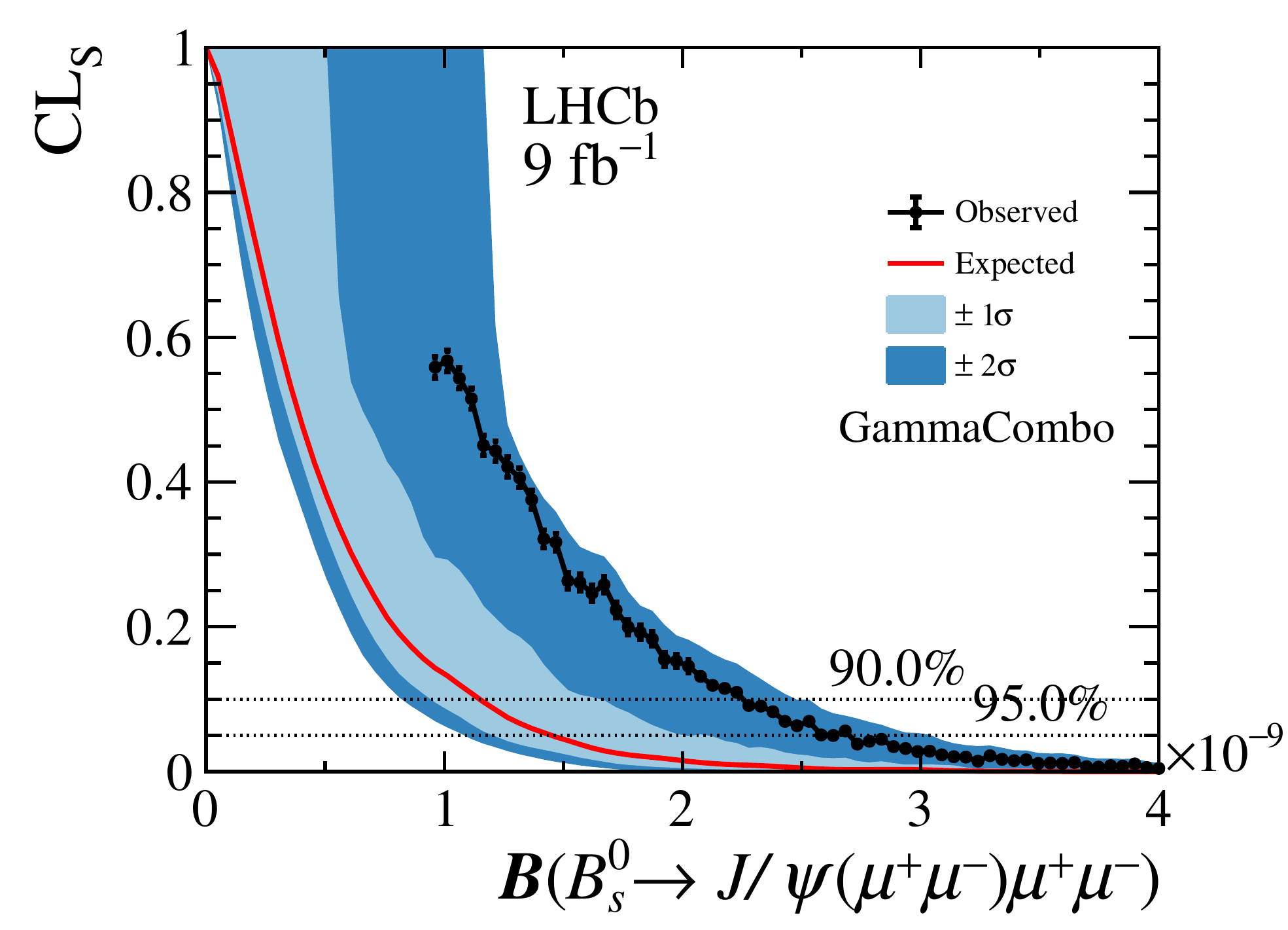}\hfill
     \includegraphics[width=0.5\linewidth]{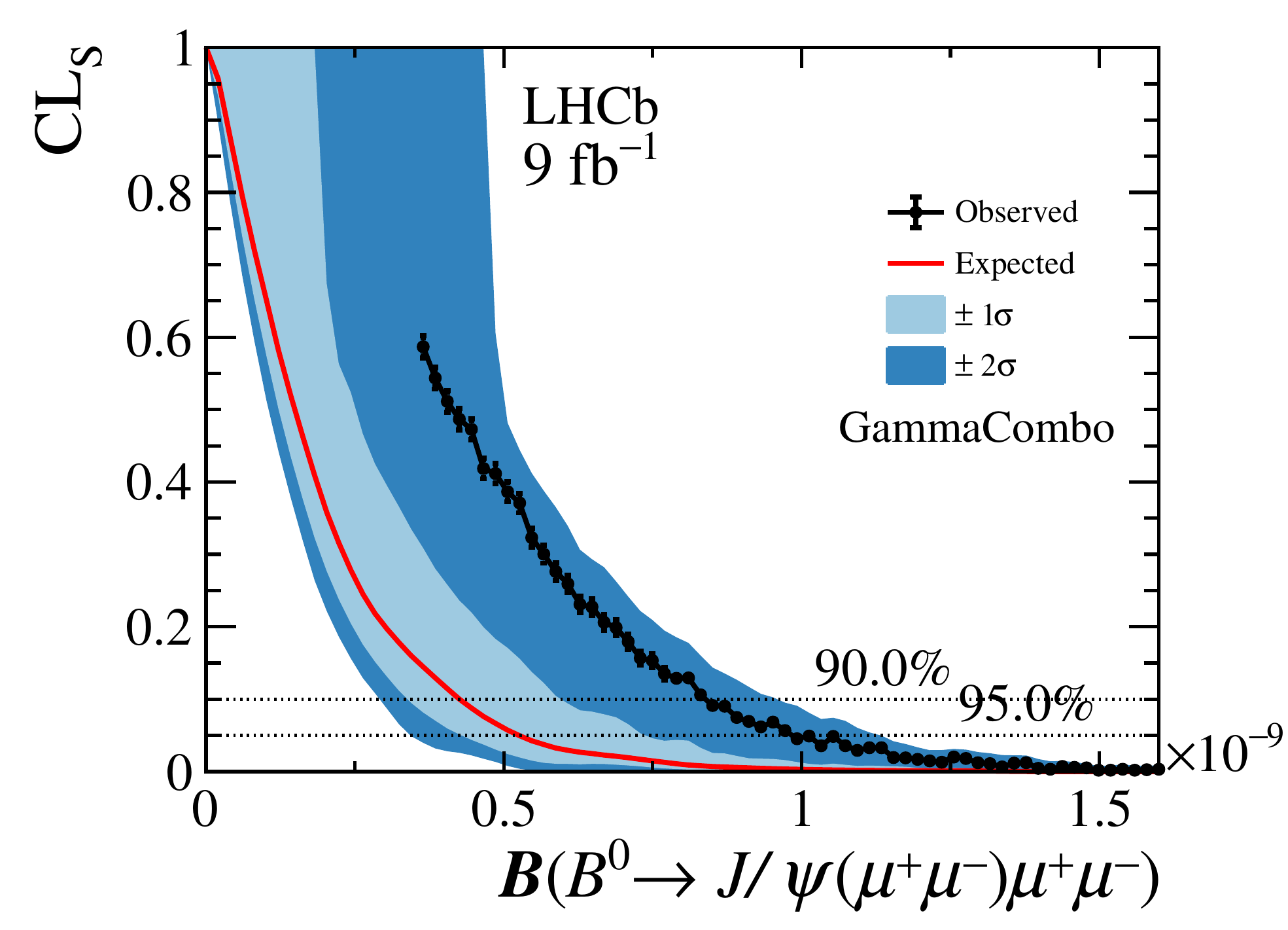}\hfill
\caption{\small Confidence levels ($\mathrm{CL_s}$) obtained from the $\text{CL}_\text{s}$ method to set limits on the branching fractions of (top left) \mbox{\BsTommmm}, (top right) \mbox{\BdTommmm}, (middle left) \mbox{\BsToaammmm}, (middle right) \mbox{\BdToaammmm}, (bottom left) \mbox{\BsTommJpsimm} and (bottom right) \mbox{\BdTommJpsimm} decays.}
\label{fig:CLs}
\end{figure}

The invariant mass distributions of \Bs and \Bz candidates passing the \BdsTommmm, \BdsToaammmm and \BdsTommJpsimm selections in the most sensitive BDT intervals are shown in Figs.~\ref{fig:BdsTommmmMassFit},~\ref{fig:BdsToaammmmMassFit} and~\ref{fig:BdsTommJpsimmMassFit}, respectively. The projections of the fits used to search for the signal decays and measure their branching fractions are overlaid. The distributions and fit projections in the remaining BDT intervals are shown in Appendix \ref{appendix}. No evidence for any of the six signal decay modes is found, with the most significant excesses found in the \BsTommJpsimm and \BdTommJpsimm searches, amounting to two standard deviations in both cases, calculated using Wilks' theorem~\cite{Wilks:1938dza}. Limits are set on the branching fractions using the $\text{CL}_\text{s}$ method~\cite{CLs} as implemented in the \textsc{GammaCombo} package~\cite{GammaCombo,LHCb-PAPER-2016-032} using a one-sided test statistic, with 80 scan points and 2000 pseudoexperiments per scan point. The limits at 95\% confidence level are
  \begin{align*}
     &\mathcal{B}\left(\BsTommmm\right) &<\, & 8.6\times 10^{-10}\,,  \\
     &\mathcal{B}\left(\BdTommmm\right) &<\, & 1.8\times 10^{-10}\,,\\
     &\mathcal{B}\left(\BsToaammmm\right) &<\, & 5.8\times 10^{-10}\,,  \\
     &\mathcal{B}\left(\BdToaammmm\right) &<\, & 2.3\times 10^{-10}\,, \\ 
     &\mathcal{B}\left(\BsTommJpsimm\right) &<\, & 2.6\times 10^{-9}\,,  \\
     &\mathcal{B}\left(\BdTommJpsimm\right) &<\, & 1.0\times 10^{-9}\,.
  \end{align*}
The corresponding $\text{CL}_\text{s}$ scans for all six decays are shown in Fig.~\ref{fig:CLs}. Note that in the case of \mbox{\BdsToaammmm} decays, these limits are evaluated assuming a promptly decaying intermediate scalar with a mass of 1\gevcc. The limits quoted for \BdsTommJpsimm decays include the \JpsiTomm branching fraction. These results constitute the most stringent limits on each of the six \BdsTommmm decays to date and supersede previous results by the LHCb experiment \cite{LHCb-PAPER-2016-043}.

\FloatBarrier

\section*{Acknowledgements}

\noindent We express our gratitude to our colleagues in the CERN
accelerator departments for the excellent performance of the LHC. We
thank the technical and administrative staff at the LHCb
institutes.
We acknowledge support from CERN and from the national agencies:
CAPES, CNPq, FAPERJ and FINEP (Brazil); 
MOST and NSFC (China); 
CNRS/IN2P3 (France); 
BMBF, DFG and MPG (Germany); 
INFN (Italy); 
NWO (Netherlands); 
MNiSW and NCN (Poland); 
MEN/IFA (Romania); 
MSHE (Russia); 
MICINN (Spain); 
SNSF and SER (Switzerland); 
NASU (Ukraine); 
STFC (United Kingdom); 
DOE NP and NSF (USA).
We acknowledge the computing resources that are provided by CERN, IN2P3
(France), KIT and DESY (Germany), INFN (Italy), SURF (Netherlands),
PIC (Spain), GridPP (United Kingdom), RRCKI and Yandex
LLC (Russia), CSCS (Switzerland), IFIN-HH (Romania), CBPF (Brazil),
PL-GRID (Poland) and NERSC (USA).
We are indebted to the communities behind the multiple open-source
software packages on which we depend.
Individual groups or members have received support from
ARC and ARDC (Australia);
AvH Foundation (Germany);
EPLANET, Marie Sk\l{}odowska-Curie Actions and ERC (European Union);
A*MIDEX, ANR, IPhU and Labex P2IO, and R\'{e}gion Auvergne-Rh\^{o}ne-Alpes (France);
Key Research Program of Frontier Sciences of CAS, CAS PIFI, CAS CCEPP, 
Fundamental Research Funds for the Central Universities, 
and Sci. \& Tech. Program of Guangzhou (China);

RFBR, RSF and Yandex LLC (Russia);
GVA, XuntaGal and GENCAT (Spain);
the Leverhulme Trust, the Royal Society
 and UKRI (United Kingdom).

\clearpage
\appendix
\section{Additional figures}
\label{appendix}

This appendix contains the distributions of $\mumu\mumu$ invariant mass for candidates passing the \BdsTommmm and \BdsTommJpsimm selections, in the lower three intervals of the BDT classifier response, with the results of the maximum-likelihood fits overlaid.

\begin{figure}[ht]
    \centering
     \includegraphics[width=0.5\linewidth]{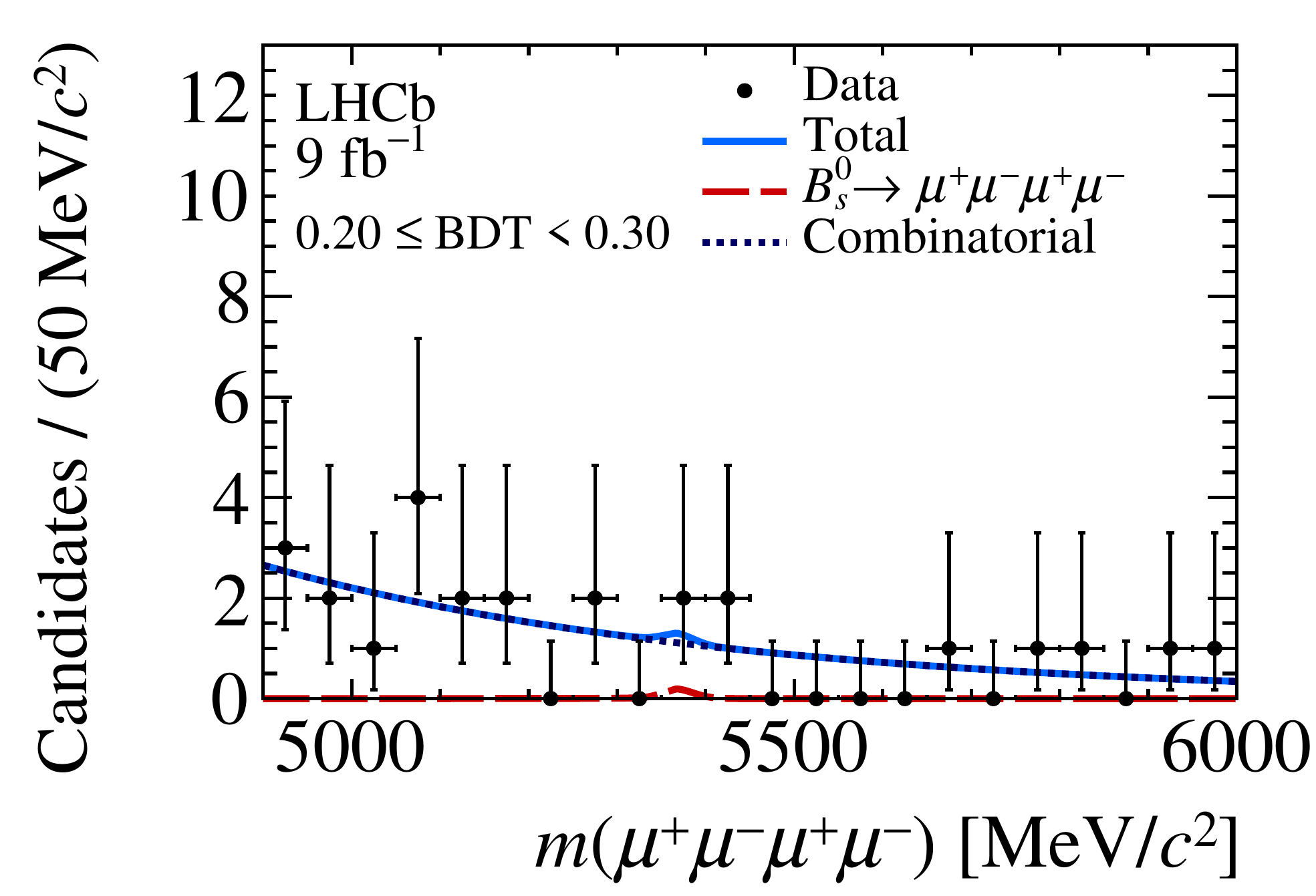}\hfill
     \includegraphics[width=0.5\linewidth]{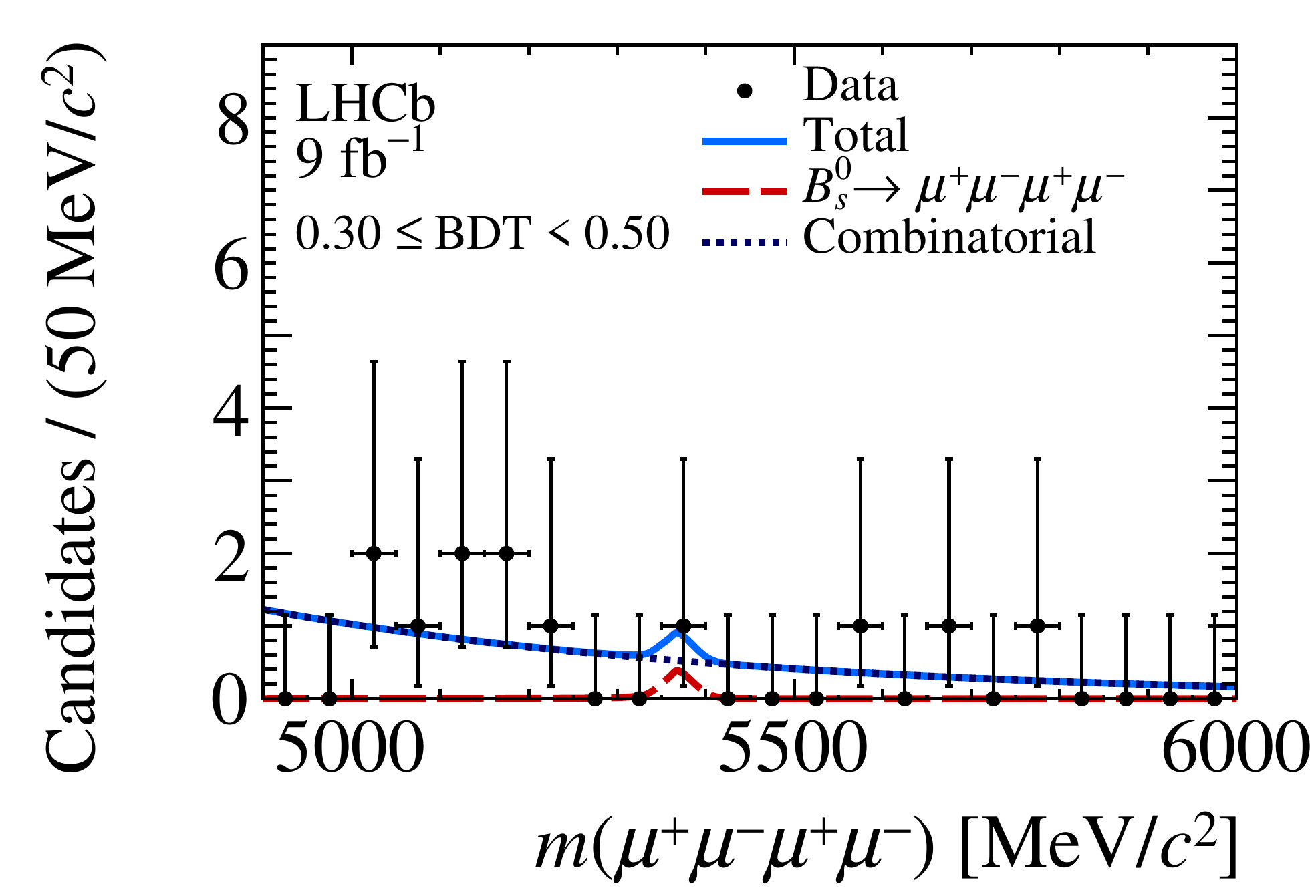}
     \includegraphics[width=0.5\linewidth]{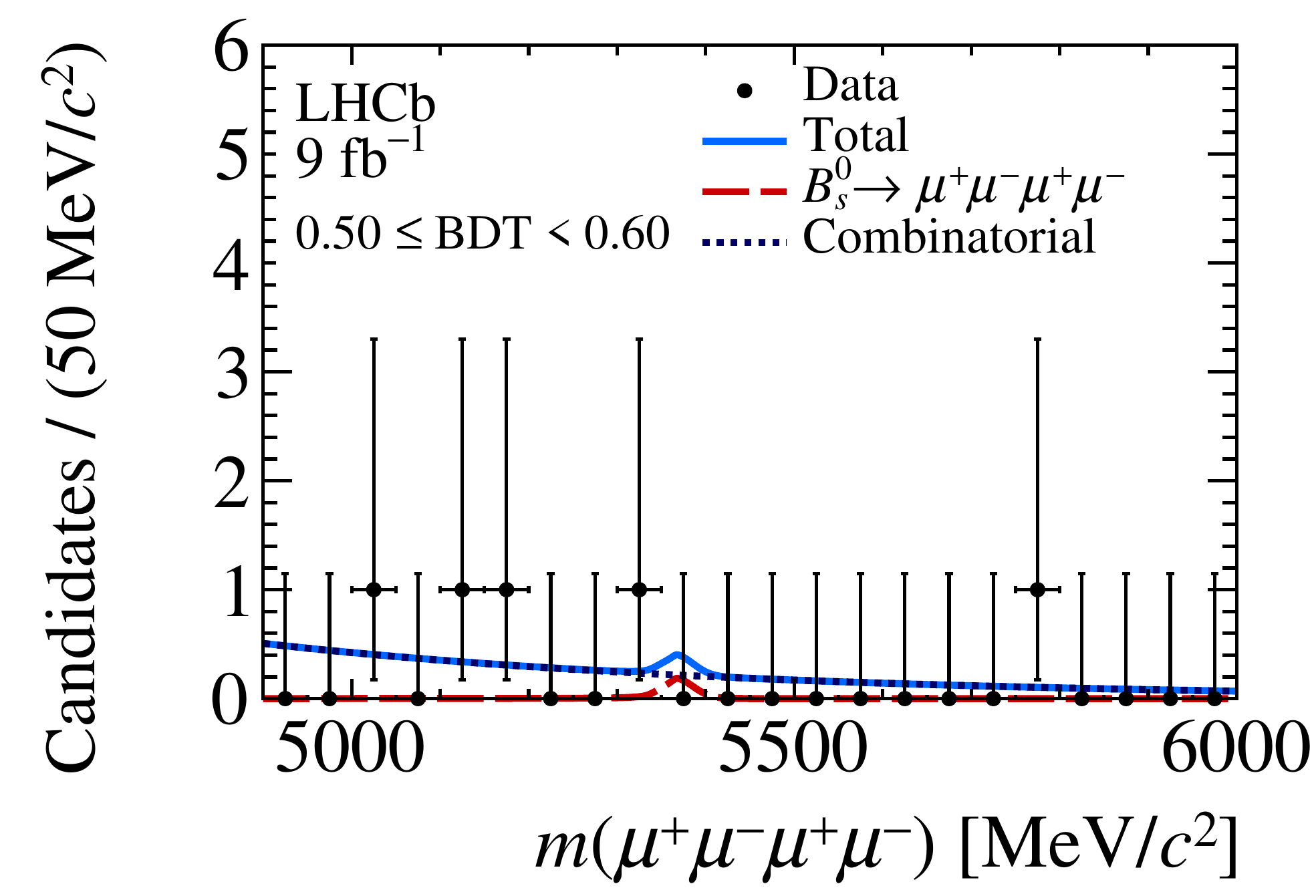}\hfill
\caption{\small 
Distribution of the $\mumu\mumu$ invariant mass of candidates passing the \mbox{\BdsTommmm} selection in (top left) the lowest BDT interval,  (top right) the second lowest BDT interval and (bottom) the second highest BDT interval, with the fit models used to determine the branching fraction of \mbox{\BsTommmm} overlaid.}
\label{fig:supp_bs24mu}
\end{figure}

\begin{figure}[ht]
    \centering
     \includegraphics[width=0.5\linewidth]{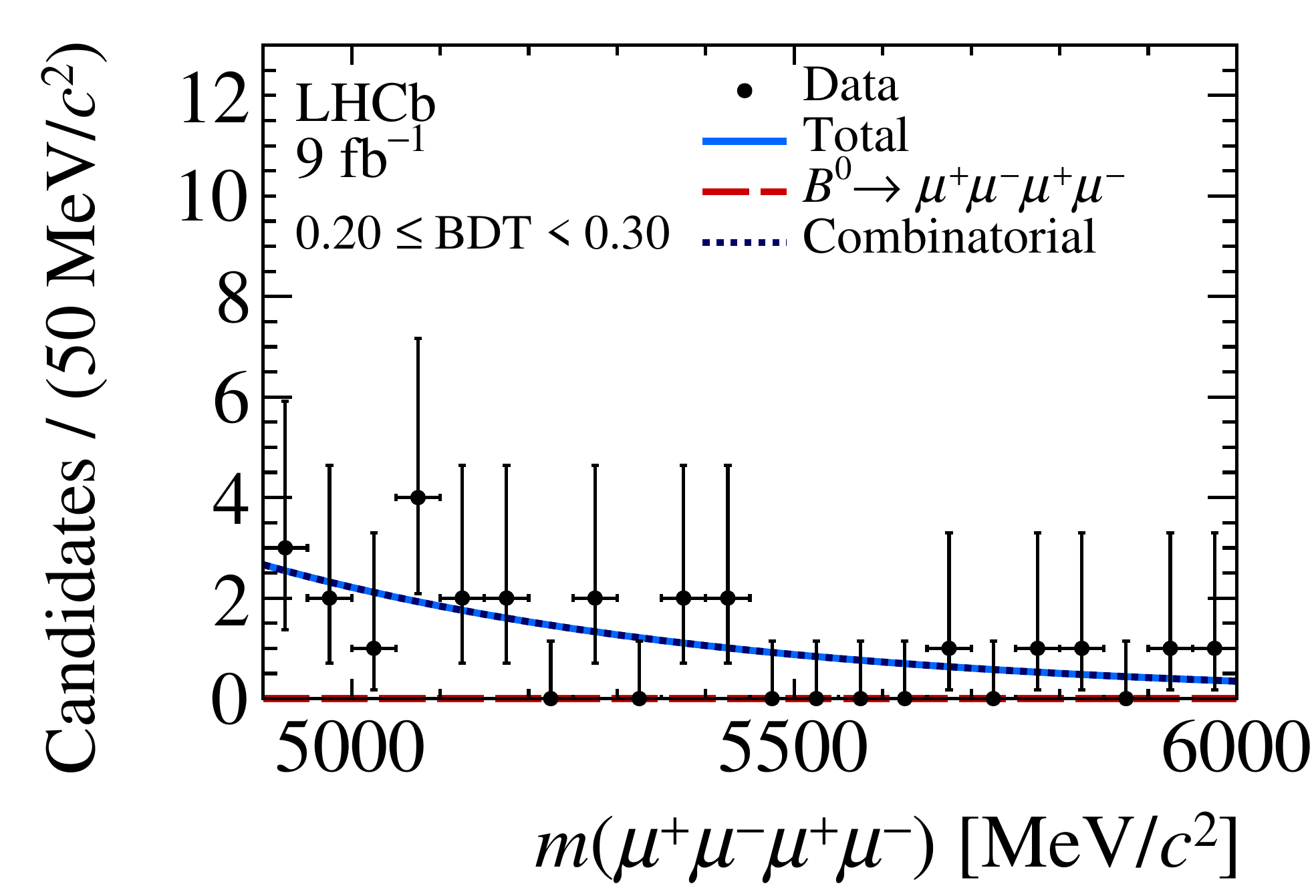}\hfill
     \includegraphics[width=0.5\linewidth]{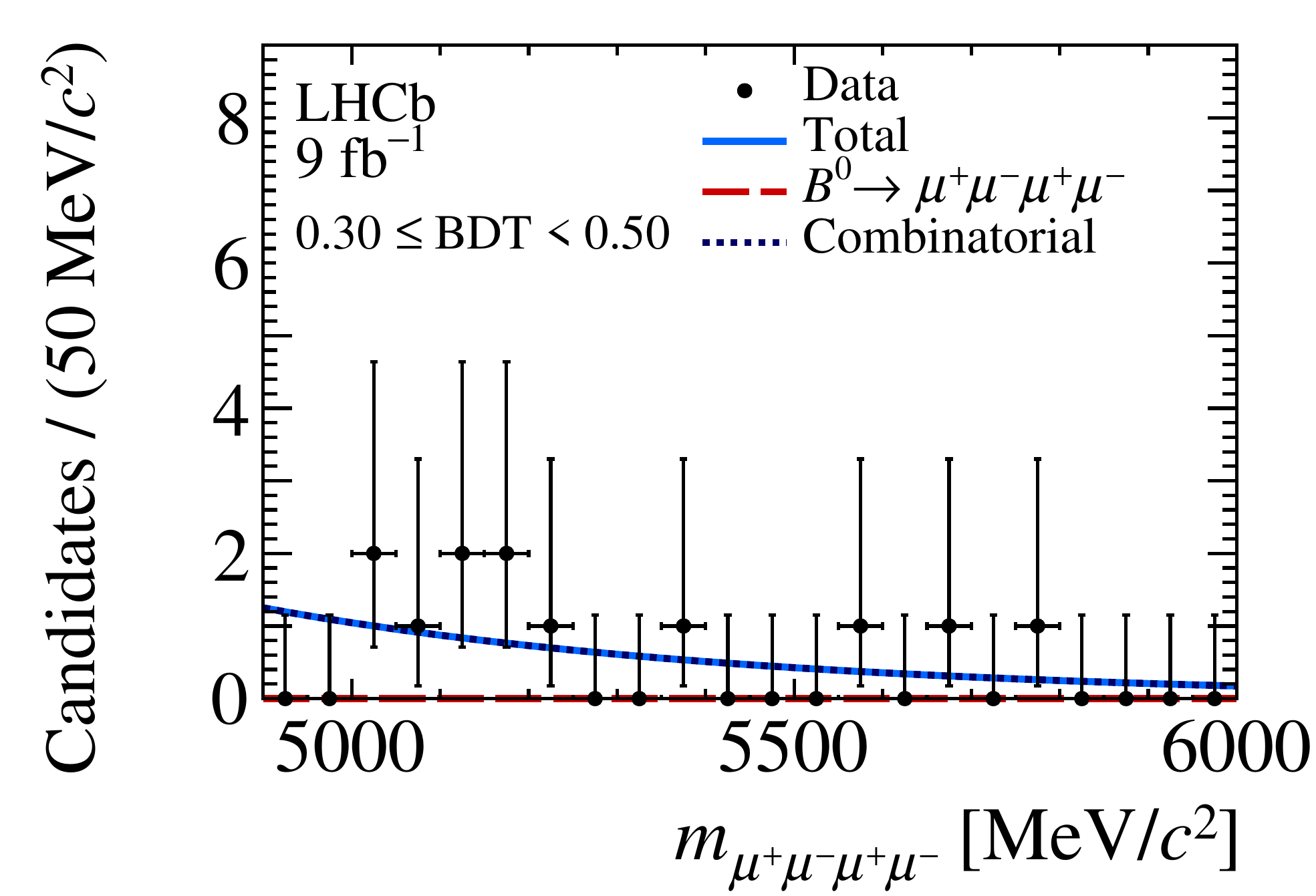}
     \includegraphics[width=0.5\linewidth]{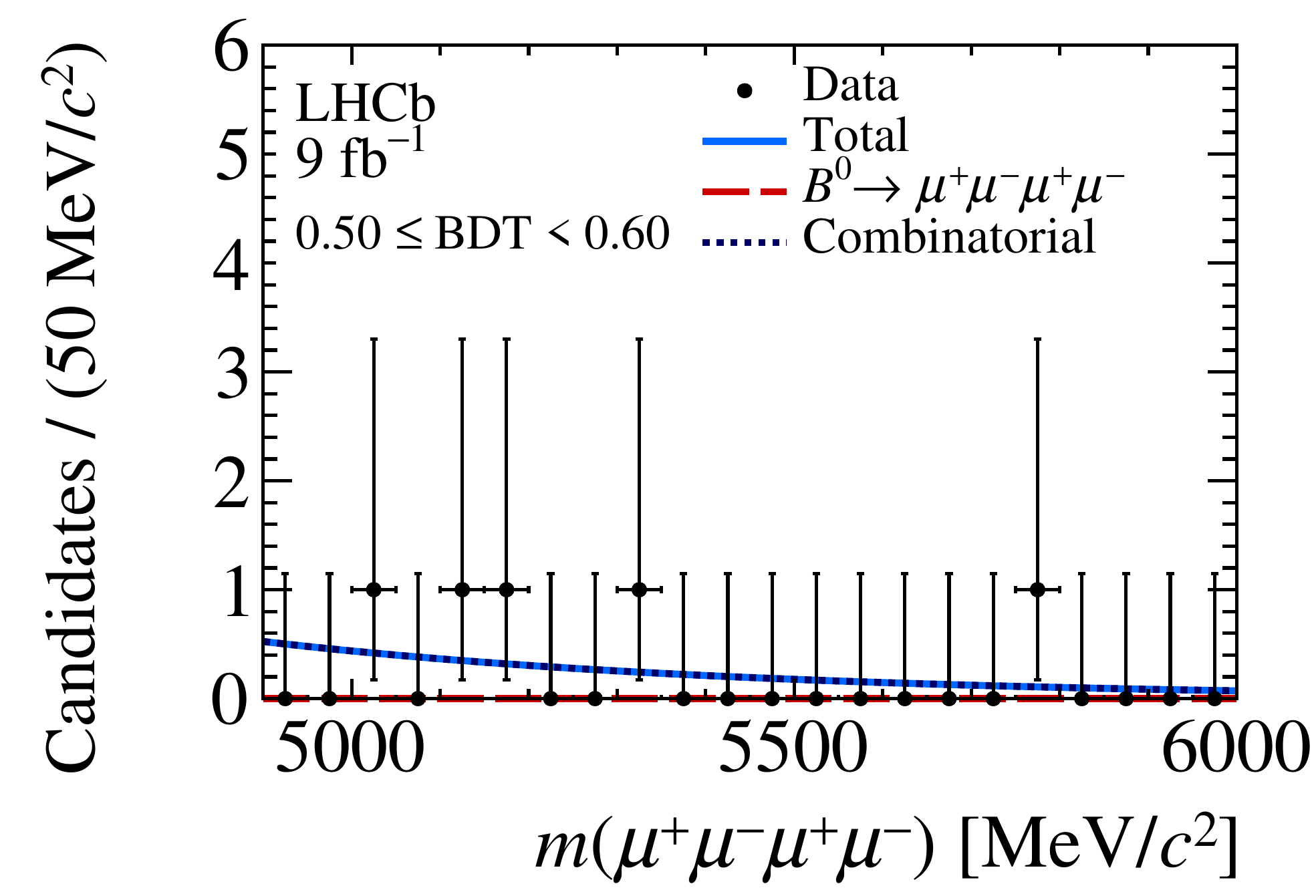}\hfill
\caption{\small Distribution of the $\mumu\mumu$ invariant mass of candidates passing the \mbox{\BdsTommmm} selection in (top left) the lowest BDT interval,  (top right) the second lowest BDT interval and (bottom) the second highest BDT interval, with the fit models used to determine the branching fraction of \mbox{\BdTommmm} overlaid.}\label{fig:supp_bd24mu}
\end{figure}

\begin{figure}[ht]
    \centering
     \includegraphics[width=0.5\linewidth]{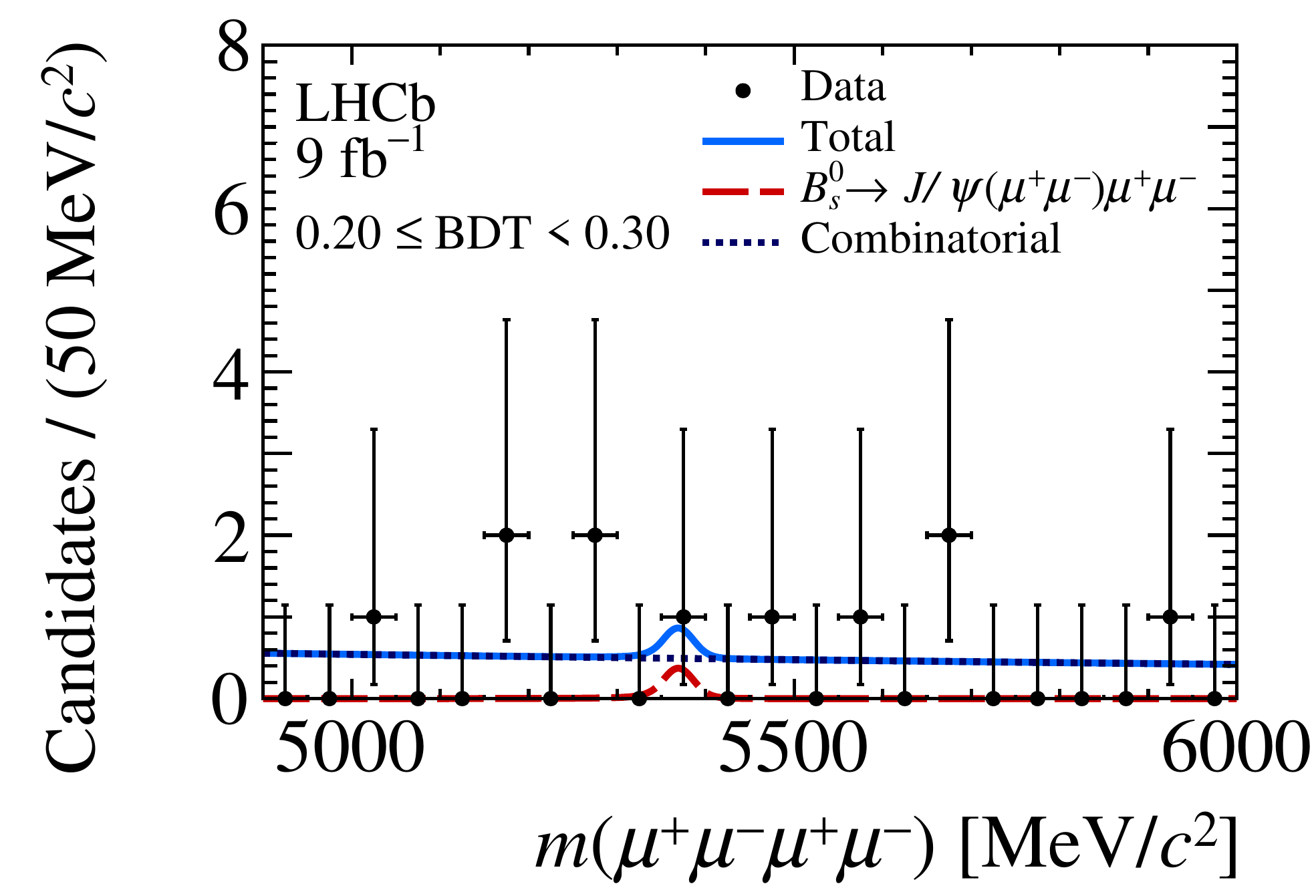}\hfill
     \includegraphics[width=0.5\linewidth]{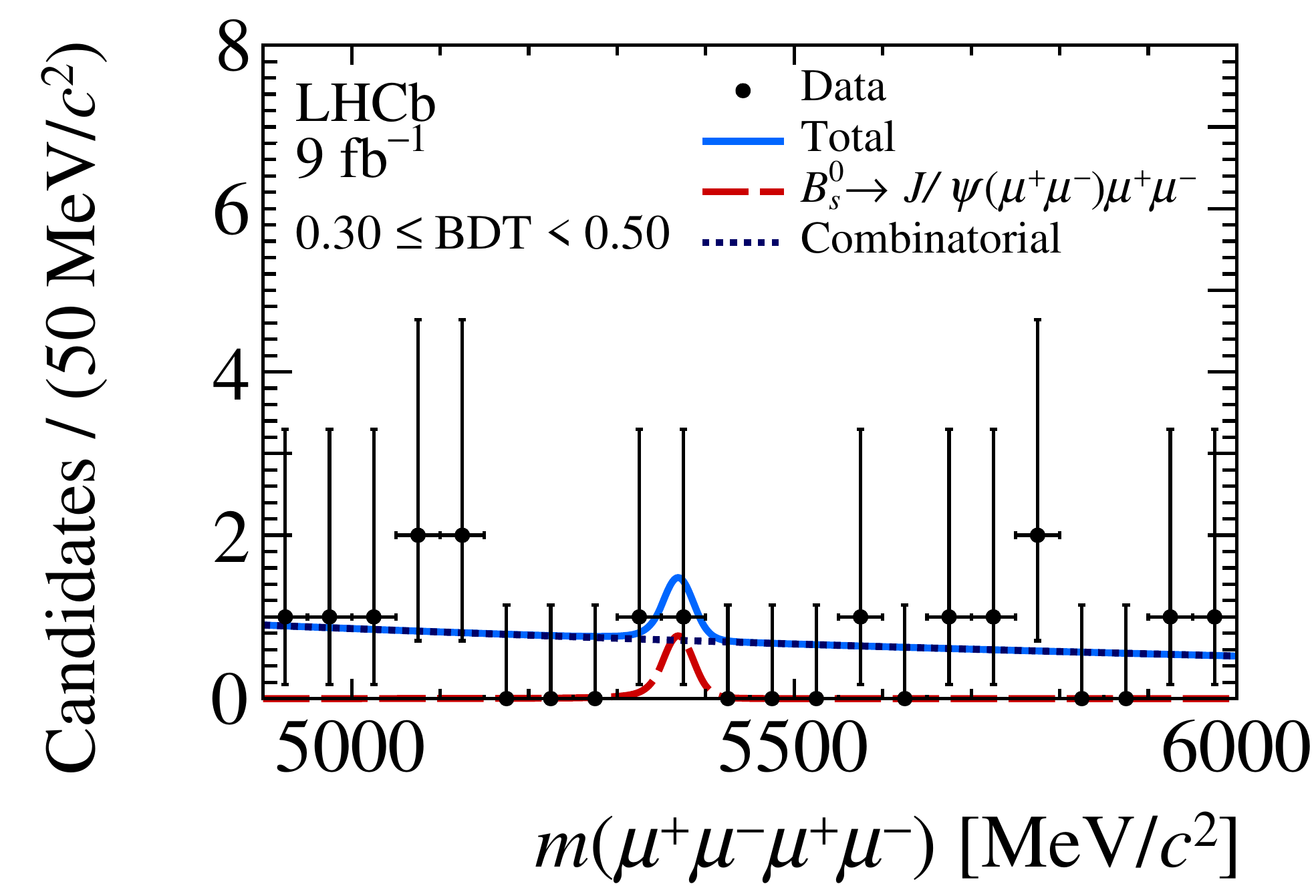}
     \includegraphics[width=0.5\linewidth]{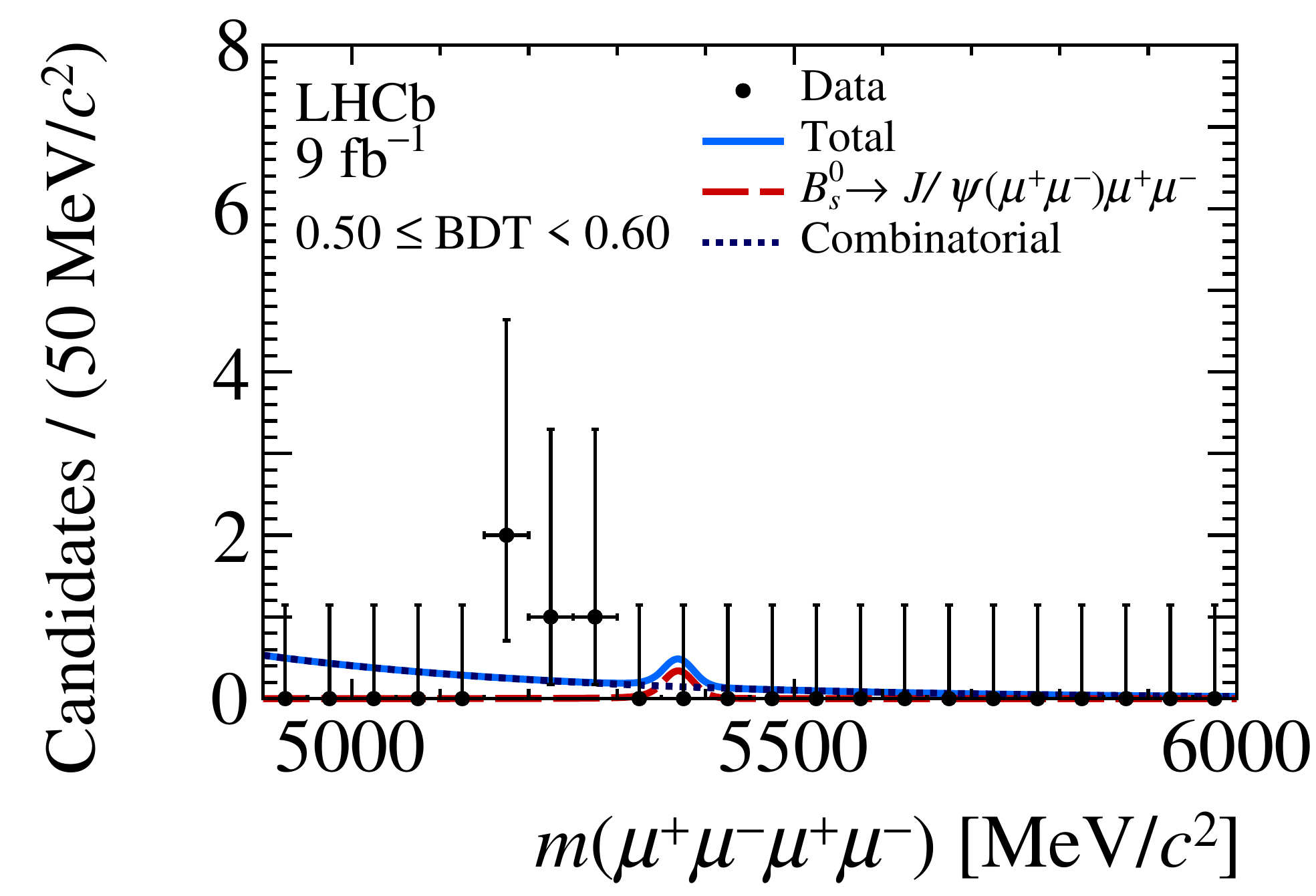}\hfill
\caption{\small Distribution of the $\mumu\mumu$ invariant mass of candidates passing the \mbox{\BdsTommJpsimm} selection in (top left) the lowest BDT interval,  (top right) the second lowest BDT interval and (bottom) the second highest BDT interval, with the fit models used to determine the branching fraction of \mbox{\BsTommJpsimm} overlaid.}\label{fig:supp_bs2mmjpsi}
\end{figure}

\begin{figure}[ht]
    \centering
     \includegraphics[width=0.5\linewidth]{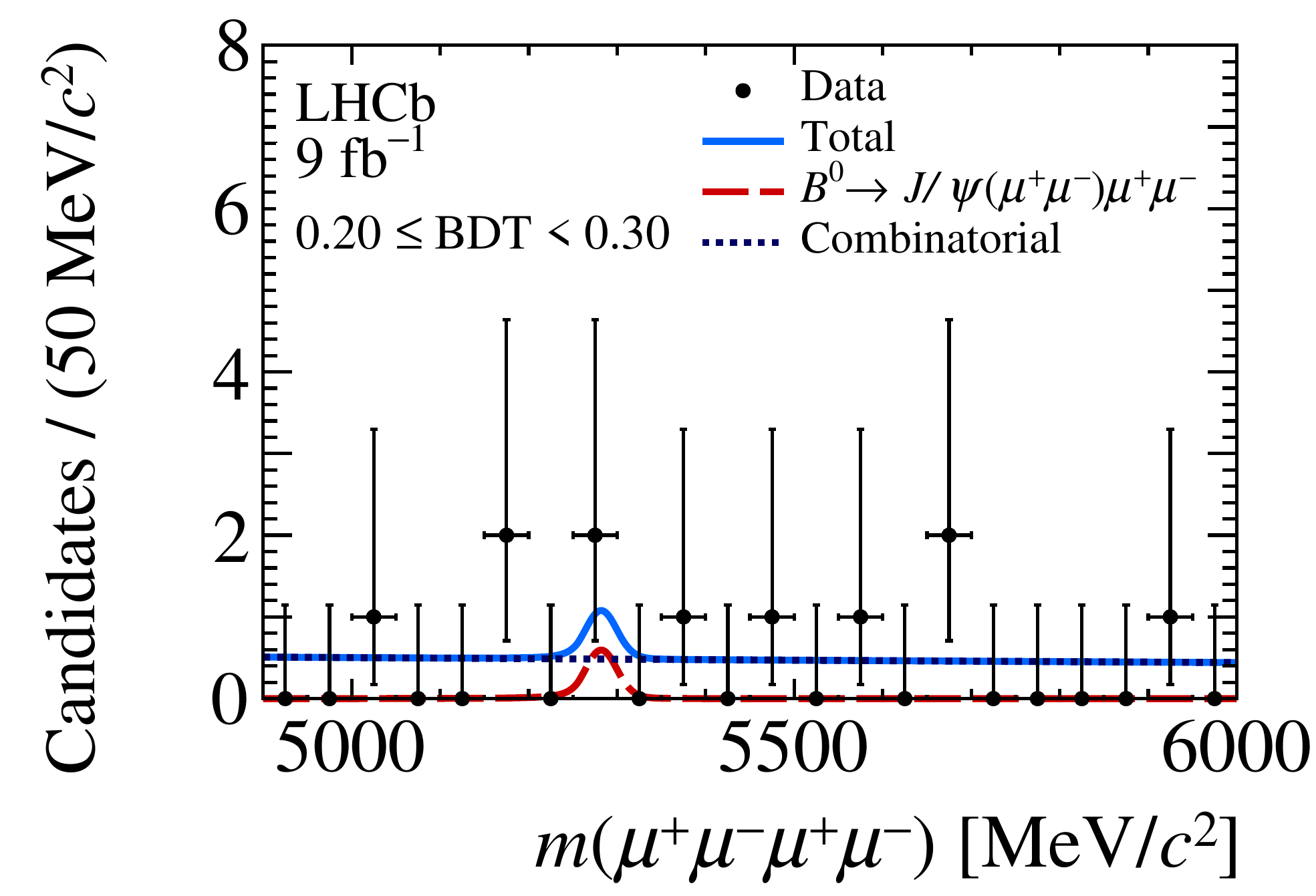}\hfill
     \includegraphics[width=0.5\linewidth]{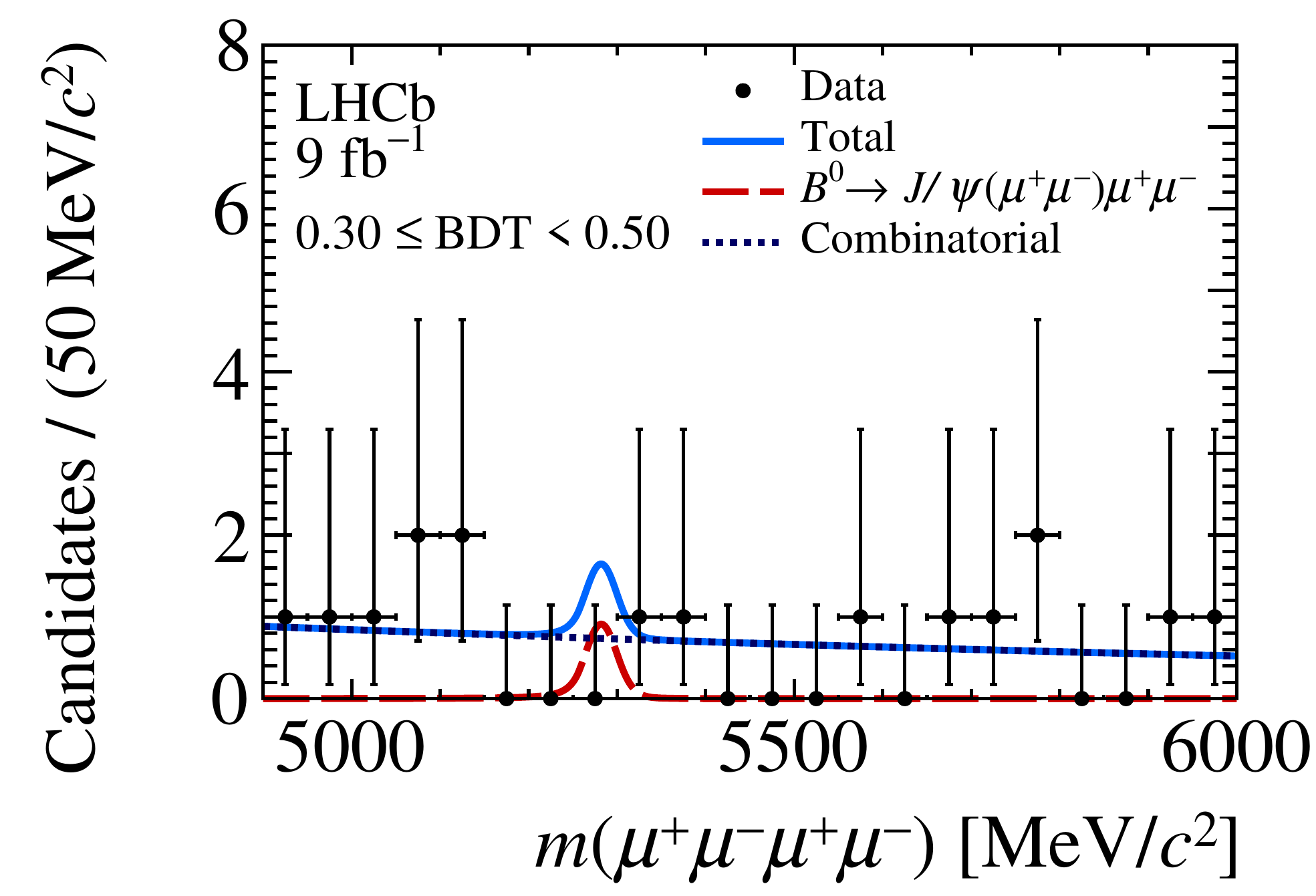}
     \includegraphics[width=0.5\linewidth]{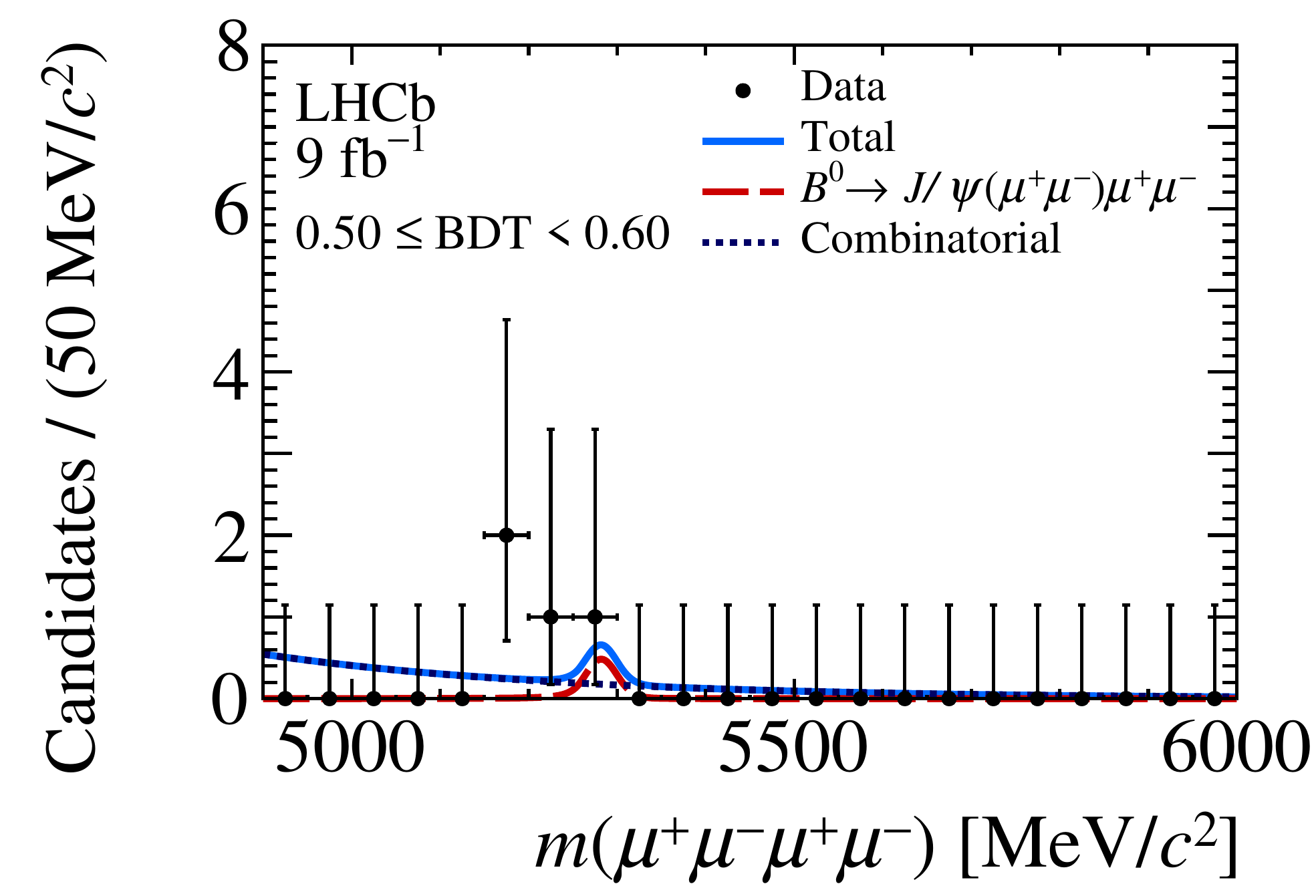}\hfill
\caption{\small Distribution of the $\mumu\mumu$ invariant mass of candidates passing the \mbox{\BdsTommJpsimm} selection in (top left) the lowest BDT interval,  (top right) the second lowest BDT interval and (bottom) the second highest BDT interval, with the fit models used to determine the branching fraction of \mbox{\BdTommJpsimm} overlaid.}\label{fig:supp_bd2mmjpsi}
\end{figure}

\clearpage

\addcontentsline{toc}{section}{References}
\bibliographystyle{LHCb}
\bibliography{main,standard,LHCb-PAPER,LHCb-CONF,LHCb-DP,LHCb-TDR}

\ifx\mcitethebibliography\mciteundefinedmacro
\PackageError{LHCb.bst}{mciteplus.sty has not been loaded}
{This bibstyle requires the use of the mciteplus package.}\fi
\providecommand{\href}[2]{#2}
\begin{mcitethebibliography}{10}
\mciteSetBstSublistMode{n}
\mciteSetBstMaxWidthForm{subitem}{\alph{mcitesubitemcount})}
\mciteSetBstSublistLabelBeginEnd{\mcitemaxwidthsubitemform\space}
{\relax}{\relax}

\bibitem{Danilina:2018uzr}
A.~V. Danilina and N.~V. Nikitin,
  \ifthenelse{\boolean{articletitles}}{\emph{{Four-leptonic decays of charged
  and neutral $B$ mesons within the Standard Model}},
  }{}\href{https://doi.org/10.1134/S1063778818030092}{Phys.\ Atom.\ Nucl.\
  \textbf{81} (2018) 347}\relax
\mciteBstWouldAddEndPuncttrue
\mciteSetBstMidEndSepPunct{\mcitedefaultmidpunct}
{\mcitedefaultendpunct}{\mcitedefaultseppunct}\relax
\EndOfBibitem
\bibitem{Demidov:2011rd}
S.~V. Demidov and D.~S. Gorbunov,
  \ifthenelse{\boolean{articletitles}}{\emph{{Flavor violating processes with
  sgoldstino pair production}},
  }{}\href{https://doi.org/10.1103/PhysRevD.85.077701}{Phys.\ Rev.\
  \textbf{D85} (2012) 077701},
  \href{http://arxiv.org/abs/1112.5230}{{\normalfont\ttfamily
  arXiv:1112.5230}}\relax
\mciteBstWouldAddEndPuncttrue
\mciteSetBstMidEndSepPunct{\mcitedefaultmidpunct}
{\mcitedefaultendpunct}{\mcitedefaultseppunct}\relax
\EndOfBibitem
\bibitem{Bauer:2017nlg}
M.~Bauer, M.~Neubert, and A.~Thamm,
  \ifthenelse{\boolean{articletitles}}{\emph{{LHC as an axion factory: probing
  an axion explanation for $(g-2)_\mu$ with exotic Higgs decays}},
  }{}\href{https://doi.org/10.1103/PhysRevLett.119.031802}{Phys.\ Rev.\ Lett.\
  \textbf{119} (2017) 031802},
  \href{http://arxiv.org/abs/1704.08207}{{\normalfont\ttfamily
  arXiv:1704.08207}}\relax
\mciteBstWouldAddEndPuncttrue
\mciteSetBstMidEndSepPunct{\mcitedefaultmidpunct}
{\mcitedefaultendpunct}{\mcitedefaultseppunct}\relax
\EndOfBibitem
\bibitem{Liu:2018xkx}
J.~Liu, C.~E.~M. Wagner, and X.-P. Wang,
  \ifthenelse{\boolean{articletitles}}{\emph{{A light complex scalar for the
  electron and muon anomalous magnetic moments}},
  }{}\href{https://doi.org/10.1007/JHEP03(2019)008}{JHEP \textbf{03} (2019)
  008}, \href{http://arxiv.org/abs/1810.11028}{{\normalfont\ttfamily
  arXiv:1810.11028}}\relax
\mciteBstWouldAddEndPuncttrue
\mciteSetBstMidEndSepPunct{\mcitedefaultmidpunct}
{\mcitedefaultendpunct}{\mcitedefaultseppunct}\relax
\EndOfBibitem
\bibitem{Aoyama:2020ynm}
T.~Aoyama {\em et~al.}, \ifthenelse{\boolean{articletitles}}{\emph{{The
  anomalous magnetic moment of the muon in the Standard Model}},
  }{}\href{https://doi.org/10.1016/j.physrep.2020.07.006}{Phys.\ Rept.\
  \textbf{887} (2020) 1},
  \href{http://arxiv.org/abs/2006.04822}{{\normalfont\ttfamily
  arXiv:2006.04822}}\relax
\mciteBstWouldAddEndPuncttrue
\mciteSetBstMidEndSepPunct{\mcitedefaultmidpunct}
{\mcitedefaultendpunct}{\mcitedefaultseppunct}\relax
\EndOfBibitem
\bibitem{Muong-2:2021ojo}
Muon g--2 collaboration, B.~Abi {\em et~al.},
  \ifthenelse{\boolean{articletitles}}{\emph{{Measurement of the positive muon
  anomalous magnetic moment to 0.46 ppm}},
  }{}\href{https://doi.org/10.1103/PhysRevLett.126.141801}{Phys.\ Rev.\ Lett.\
  \textbf{126} (2021) 141801},
  \href{http://arxiv.org/abs/2104.03281}{{\normalfont\ttfamily
  arXiv:2104.03281}}\relax
\mciteBstWouldAddEndPuncttrue
\mciteSetBstMidEndSepPunct{\mcitedefaultmidpunct}
{\mcitedefaultendpunct}{\mcitedefaultseppunct}\relax
\EndOfBibitem
\bibitem{Muong-2:2006rrc}
Muon g--2 collaboration, G.~W. Bennett {\em et~al.},
  \ifthenelse{\boolean{articletitles}}{\emph{{Final report of the muon E821
  anomalous magnetic moment measurement at BNL}},
  }{}\href{https://doi.org/10.1103/PhysRevD.73.072003}{Phys.\ Rev.\
  \textbf{D73} (2006) 072003},
  \href{http://arxiv.org/abs/hep-ex/0602035}{{\normalfont\ttfamily
  arXiv:hep-ex/0602035}}\relax
\mciteBstWouldAddEndPuncttrue
\mciteSetBstMidEndSepPunct{\mcitedefaultmidpunct}
{\mcitedefaultendpunct}{\mcitedefaultseppunct}\relax
\EndOfBibitem
\bibitem{LHCb-PAPER-2021-038}
LHCb collaboration, R.~Aaij {\em et~al.},
  \ifthenelse{\boolean{articletitles}}{\emph{{Tests of lepton universality
  using $B^0 \to K^0_S \ell^+\ell^-$ and $B^+\to K^{*+}\ell^+ \ell^-$ decays}},
  }{}\href{http://arxiv.org/abs/2110.09501}{{\normalfont\ttfamily
  arXiv:2110.09501}}, {submitted to PRL}\relax
\mciteBstWouldAddEndPuncttrue
\mciteSetBstMidEndSepPunct{\mcitedefaultmidpunct}
{\mcitedefaultendpunct}{\mcitedefaultseppunct}\relax
\EndOfBibitem
\bibitem{LHCb-PAPER-2021-004}
LHCb collaboration, R.~Aaij {\em et~al.},
  \ifthenelse{\boolean{articletitles}}{\emph{{Test of lepton universality in
  beauty-quark decays}},
  }{}\href{http://arxiv.org/abs/2103.11769}{{\normalfont\ttfamily
  arXiv:2103.11769}}, {to appear in Nature Physics}\relax
\mciteBstWouldAddEndPuncttrue
\mciteSetBstMidEndSepPunct{\mcitedefaultmidpunct}
{\mcitedefaultendpunct}{\mcitedefaultseppunct}\relax
\EndOfBibitem
\bibitem{LHCb-PAPER-2019-040}
LHCb collaboration, R.~Aaij {\em et~al.},
  \ifthenelse{\boolean{articletitles}}{\emph{{Test of lepton universality using
  \mbox{\decay{\Lb}{p\Km\ellell}} decays}},
  }{}\href{https://doi.org/10.1007/JHEP05(2020)040}{JHEP \textbf{05} (2020)
  040}, \href{http://arxiv.org/abs/1912.08139}{{\normalfont\ttfamily
  arXiv:1912.08139}}\relax
\mciteBstWouldAddEndPuncttrue
\mciteSetBstMidEndSepPunct{\mcitedefaultmidpunct}
{\mcitedefaultendpunct}{\mcitedefaultseppunct}\relax
\EndOfBibitem
\bibitem{LHCb-PAPER-2017-013}
LHCb collaboration, R.~Aaij {\em et~al.},
  \ifthenelse{\boolean{articletitles}}{\emph{{Test of lepton universality with
  \mbox{\decay{\Bz}{\Kstarz\ellell}} decays}},
  }{}\href{https://doi.org/10.1007/JHEP08(2017)055}{JHEP \textbf{08} (2017)
  055}, \href{http://arxiv.org/abs/1705.05802}{{\normalfont\ttfamily
  arXiv:1705.05802}}\relax
\mciteBstWouldAddEndPuncttrue
\mciteSetBstMidEndSepPunct{\mcitedefaultmidpunct}
{\mcitedefaultendpunct}{\mcitedefaultseppunct}\relax
\EndOfBibitem
\bibitem{LHCb-PAPER-2021-022}
LHCb collaboration, R.~Aaij {\em et~al.},
  \ifthenelse{\boolean{articletitles}}{\emph{{Angular analysis of the rare
  decay $\Bs \to \Pphi \mup \mun$}},
  }{}\href{https://doi.org/10.1007/JHEP11(2021)043}{JHEP \textbf{11} (2021)
  043}, \href{http://arxiv.org/abs/2107.13428}{{\normalfont\ttfamily
  arXiv:2107.13428}}\relax
\mciteBstWouldAddEndPuncttrue
\mciteSetBstMidEndSepPunct{\mcitedefaultmidpunct}
{\mcitedefaultendpunct}{\mcitedefaultseppunct}\relax
\EndOfBibitem
\bibitem{LHCb-PAPER-2020-041}
LHCb collaboration, R.~Aaij {\em et~al.},
  \ifthenelse{\boolean{articletitles}}{\emph{{Angular analysis of the $B^{+}\to
  K^{\ast+}\mu^+\mu^-$ decay}},
  }{}\href{https://doi.org/10.1103/PhysRevLett.126.161802}{Phys.\ Rev.\ Lett.\
  \textbf{126} (2021) 161802},
  \href{http://arxiv.org/abs/2012.13241}{{\normalfont\ttfamily
  arXiv:2012.13241}}\relax
\mciteBstWouldAddEndPuncttrue
\mciteSetBstMidEndSepPunct{\mcitedefaultmidpunct}
{\mcitedefaultendpunct}{\mcitedefaultseppunct}\relax
\EndOfBibitem
\bibitem{LHCb-PAPER-2020-002}
LHCb collaboration, R.~Aaij {\em et~al.},
  \ifthenelse{\boolean{articletitles}}{\emph{{Measurement of \CP-averaged
  observables in the \mbox{\decay{\Bz}{\Kstarz\mumu}} decay}},
  }{}\href{https://doi.org/10.1103/PhysRevLett.125.011802}{Phys.\ Rev.\ Lett.\
  \textbf{125} (2020) 011802},
  \href{http://arxiv.org/abs/2003.04831}{{\normalfont\ttfamily
  arXiv:2003.04831}}\relax
\mciteBstWouldAddEndPuncttrue
\mciteSetBstMidEndSepPunct{\mcitedefaultmidpunct}
{\mcitedefaultendpunct}{\mcitedefaultseppunct}\relax
\EndOfBibitem
\bibitem{Chala:2019vzu}
M.~Chala, U.~Egede, and M.~Spannowsky,
  \ifthenelse{\boolean{articletitles}}{\emph{{Searching new physics in rare
  $B$-meson decays into multiple muons}},
  }{}\href{https://doi.org/10.1140/epjc/s10052-019-6946-6}{Eur.\ Phys.\ J.\
  \textbf{C79} (2019) 431},
  \href{http://arxiv.org/abs/1902.10156}{{\normalfont\ttfamily
  arXiv:1902.10156}}\relax
\mciteBstWouldAddEndPuncttrue
\mciteSetBstMidEndSepPunct{\mcitedefaultmidpunct}
{\mcitedefaultendpunct}{\mcitedefaultseppunct}\relax
\EndOfBibitem
\bibitem{LHCb-PAPER-2016-043}
LHCb collaboration, R.~Aaij {\em et~al.},
  \ifthenelse{\boolean{articletitles}}{\emph{{Search for decays of neutral
  beauty mesons into four muons}},
  }{}\href{https://doi.org/10.1007/JHEP03(2017)001}{JHEP \textbf{03} (2017)
  001}, \href{http://arxiv.org/abs/1611.07704}{{\normalfont\ttfamily
  arXiv:1611.07704}}\relax
\mciteBstWouldAddEndPuncttrue
\mciteSetBstMidEndSepPunct{\mcitedefaultmidpunct}
{\mcitedefaultendpunct}{\mcitedefaultseppunct}\relax
\EndOfBibitem
\bibitem{Evans:1999zc}
D.~H. Evans, B.~Grinstein, and D.~R. Nolte,
  \ifthenelse{\boolean{articletitles}}{\emph{{Short distance analysis of $\Bzb
  \rightarrow \jpsi \ep \en$, $\Bzb \rightarrow \etac \ep \en$, $\Bzb
  \rightarrow \Dstarz \ep \en$ and $\Bzb \rightarrow \Dz \ep \en$}},
  }{}\href{https://doi.org/10.1016/S0550-3213(00)00159-0}{Nucl.\ Phys.\
  \textbf{B577} (2000) 240},
  \href{http://arxiv.org/abs/hep-ph/9906528}{{\normalfont\ttfamily
  arXiv:hep-ph/9906528}}\relax
\mciteBstWouldAddEndPuncttrue
\mciteSetBstMidEndSepPunct{\mcitedefaultmidpunct}
{\mcitedefaultendpunct}{\mcitedefaultseppunct}\relax
\EndOfBibitem
\bibitem{PDG2020}
Particle Data Group, P.~A. Zyla {\em et~al.},
  \ifthenelse{\boolean{articletitles}}{\emph{{\href{http://pdg.lbl.gov/}{Review
  of particle physics}}}, }{}\href{https://doi.org/10.1093/ptep/ptaa104}{Prog.\
  Theor.\ Exp.\ Phys.\  \textbf{2020} (2020) 083C01}\relax
\mciteBstWouldAddEndPuncttrue
\mciteSetBstMidEndSepPunct{\mcitedefaultmidpunct}
{\mcitedefaultendpunct}{\mcitedefaultseppunct}\relax
\EndOfBibitem
\bibitem{LHCb-PAPER-2020-046}
LHCb collaboration, R.~Aaij {\em et~al.},
  \ifthenelse{\boolean{articletitles}}{\emph{{Precise measurement of the
  $f_s/f_d$ ratio of fragmentation fractions and of $B^0_s$ decay branching
  fractions}}, }{}\href{https://doi.org/10.1103/PhysRevD.104.032005}{Phys.\
  Rev.\  \textbf{D104} (2021) 032005},
  \href{http://arxiv.org/abs/2103.06810}{{\normalfont\ttfamily
  arXiv:2103.06810}}\relax
\mciteBstWouldAddEndPuncttrue
\mciteSetBstMidEndSepPunct{\mcitedefaultmidpunct}
{\mcitedefaultendpunct}{\mcitedefaultseppunct}\relax
\EndOfBibitem
\bibitem{LHCb-DP-2008-001}
LHCb collaboration, A.~A. Alves~Jr.\ {\em et~al.},
  \ifthenelse{\boolean{articletitles}}{\emph{{The \lhcb detector at the LHC}},
  }{}\href{https://doi.org/10.1088/1748-0221/3/08/S08005}{JINST \textbf{3}
  (2008) S08005}\relax
\mciteBstWouldAddEndPuncttrue
\mciteSetBstMidEndSepPunct{\mcitedefaultmidpunct}
{\mcitedefaultendpunct}{\mcitedefaultseppunct}\relax
\EndOfBibitem
\bibitem{LHCb-DP-2014-002}
LHCb collaboration, R.~Aaij {\em et~al.},
  \ifthenelse{\boolean{articletitles}}{\emph{{LHCb detector performance}},
  }{}\href{https://doi.org/10.1142/S0217751X15300227}{Int.\ J.\ Mod.\ Phys.\
  \textbf{A30} (2015) 1530022},
  \href{http://arxiv.org/abs/1412.6352}{{\normalfont\ttfamily
  arXiv:1412.6352}}\relax
\mciteBstWouldAddEndPuncttrue
\mciteSetBstMidEndSepPunct{\mcitedefaultmidpunct}
{\mcitedefaultendpunct}{\mcitedefaultseppunct}\relax
\EndOfBibitem
\bibitem{Sjostrand:2007gs}
T.~Sj\"{o}strand, S.~Mrenna, and P.~Skands,
  \ifthenelse{\boolean{articletitles}}{\emph{{A brief introduction to PYTHIA
  8.1}}, }{}\href{https://doi.org/10.1016/j.cpc.2008.01.036}{Comput.\ Phys.\
  Commun.\  \textbf{178} (2008) 852},
  \href{http://arxiv.org/abs/0710.3820}{{\normalfont\ttfamily
  arXiv:0710.3820}}\relax
\mciteBstWouldAddEndPuncttrue
\mciteSetBstMidEndSepPunct{\mcitedefaultmidpunct}
{\mcitedefaultendpunct}{\mcitedefaultseppunct}\relax
\EndOfBibitem
\bibitem{Sjostrand:2006za}
T.~Sj\"{o}strand, S.~Mrenna, and P.~Skands,
  \ifthenelse{\boolean{articletitles}}{\emph{{PYTHIA 6.4 physics and manual}},
  }{}\href{https://doi.org/10.1088/1126-6708/2006/05/026}{JHEP \textbf{05}
  (2006) 026}, \href{http://arxiv.org/abs/hep-ph/0603175}{{\normalfont\ttfamily
  arXiv:hep-ph/0603175}}\relax
\mciteBstWouldAddEndPuncttrue
\mciteSetBstMidEndSepPunct{\mcitedefaultmidpunct}
{\mcitedefaultendpunct}{\mcitedefaultseppunct}\relax
\EndOfBibitem
\bibitem{LHCb-PROC-2010-056}
I.~Belyaev {\em et~al.}, \ifthenelse{\boolean{articletitles}}{\emph{{Handling
  of the generation of primary events in Gauss, the LHCb simulation
  framework}}, }{}\href{https://doi.org/10.1088/1742-6596/331/3/032047}{J.\
  Phys.\ Conf.\ Ser.\  \textbf{331} (2011) 032047}\relax
\mciteBstWouldAddEndPuncttrue
\mciteSetBstMidEndSepPunct{\mcitedefaultmidpunct}
{\mcitedefaultendpunct}{\mcitedefaultseppunct}\relax
\EndOfBibitem
\bibitem{Lange:2001uf}
D.~J. Lange, \ifthenelse{\boolean{articletitles}}{\emph{{The EvtGen particle
  decay simulation package}},
  }{}\href{https://doi.org/10.1016/S0168-9002(01)00089-4}{Nucl.\ Instrum.\
  Meth.\  \textbf{A462} (2001) 152}\relax
\mciteBstWouldAddEndPuncttrue
\mciteSetBstMidEndSepPunct{\mcitedefaultmidpunct}
{\mcitedefaultendpunct}{\mcitedefaultseppunct}\relax
\EndOfBibitem
\bibitem{davidson2015photos}
N.~Davidson, T.~Przedzinski, and Z.~Was,
  \ifthenelse{\boolean{articletitles}}{\emph{{PHOTOS interface in C++:
  Technical and physics documentation}},
  }{}\href{https://doi.org/https://doi.org/10.1016/j.cpc.2015.09.013}{Comp.\
  Phys.\ Comm.\  \textbf{199} (2016) 86},
  \href{http://arxiv.org/abs/1011.0937}{{\normalfont\ttfamily
  arXiv:1011.0937}}\relax
\mciteBstWouldAddEndPuncttrue
\mciteSetBstMidEndSepPunct{\mcitedefaultmidpunct}
{\mcitedefaultendpunct}{\mcitedefaultseppunct}\relax
\EndOfBibitem
\bibitem{Allison:2006ve}
Geant4 collaboration, J.~Allison {\em et~al.},
  \ifthenelse{\boolean{articletitles}}{\emph{{Geant4 developments and
  applications}}, }{}\href{https://doi.org/10.1109/TNS.2006.869826}{IEEE
  Trans.\ Nucl.\ Sci.\  \textbf{53} (2006) 270}\relax
\mciteBstWouldAddEndPuncttrue
\mciteSetBstMidEndSepPunct{\mcitedefaultmidpunct}
{\mcitedefaultendpunct}{\mcitedefaultseppunct}\relax
\EndOfBibitem
\bibitem{Agostinelli:2002hh}
Geant4 collaboration, S.~Agostinelli {\em et~al.},
  \ifthenelse{\boolean{articletitles}}{\emph{{Geant4: A simulation toolkit}},
  }{}\href{https://doi.org/10.1016/S0168-9002(03)01368-8}{Nucl.\ Instrum.\
  Meth.\  \textbf{A506} (2003) 250}\relax
\mciteBstWouldAddEndPuncttrue
\mciteSetBstMidEndSepPunct{\mcitedefaultmidpunct}
{\mcitedefaultendpunct}{\mcitedefaultseppunct}\relax
\EndOfBibitem
\bibitem{LHCb-PROC-2011-006}
M.~Clemencic {\em et~al.}, \ifthenelse{\boolean{articletitles}}{\emph{{The
  \lhcb simulation application, Gauss: Design, evolution and experience}},
  }{}\href{https://doi.org/10.1088/1742-6596/331/3/032023}{J.\ Phys.\ Conf.\
  Ser.\  \textbf{331} (2011) 032023}\relax
\mciteBstWouldAddEndPuncttrue
\mciteSetBstMidEndSepPunct{\mcitedefaultmidpunct}
{\mcitedefaultendpunct}{\mcitedefaultseppunct}\relax
\EndOfBibitem
\bibitem{Ali:1999mm}
A.~Ali, P.~Ball, L.~T. Handoko, and G.~Hiller,
  \ifthenelse{\boolean{articletitles}}{\emph{{A comparative study of the decays
  $B \to$ ($K$, $K^{*)} \ell^+ \ell^-$ in standard model and supersymmetric
  theories}}, }{}\href{https://doi.org/10.1103/PhysRevD.61.074024}{Phys.\ Rev.\
   \textbf{D61} (2000) 074024},
  \href{http://arxiv.org/abs/hep-ph/9910221}{{\normalfont\ttfamily
  arXiv:hep-ph/9910221}}\relax
\mciteBstWouldAddEndPuncttrue
\mciteSetBstMidEndSepPunct{\mcitedefaultmidpunct}
{\mcitedefaultendpunct}{\mcitedefaultseppunct}\relax
\EndOfBibitem
\bibitem{LHCb-DP-2013-001}
F.~Archilli {\em et~al.},
  \ifthenelse{\boolean{articletitles}}{\emph{{Performance of the muon
  identification at LHCb}},
  }{}\href{https://doi.org/10.1088/1748-0221/8/10/P10020}{JINST \textbf{8}
  (2013) P10020}, \href{http://arxiv.org/abs/1306.0249}{{\normalfont\ttfamily
  arXiv:1306.0249}}\relax
\mciteBstWouldAddEndPuncttrue
\mciteSetBstMidEndSepPunct{\mcitedefaultmidpunct}
{\mcitedefaultendpunct}{\mcitedefaultseppunct}\relax
\EndOfBibitem
\bibitem{Hocker:2007ht}
H.~Voss, A.~Hoecker, J.~Stelzer, and F.~Tegenfeldt,
  \ifthenelse{\boolean{articletitles}}{\emph{{TMVA - Toolkit for Multivariate
  Data Analysis with ROOT}}, }{}\href{https://doi.org/10.22323/1.050.0040}{PoS
  \textbf{ACAT} (2007) 040}\relax
\mciteBstWouldAddEndPuncttrue
\mciteSetBstMidEndSepPunct{\mcitedefaultmidpunct}
{\mcitedefaultendpunct}{\mcitedefaultseppunct}\relax
\EndOfBibitem
\bibitem{Blum:1999:BHB:307400.307439}
A.~Blum, A.~Kalai, and J.~Langford,
  \ifthenelse{\boolean{articletitles}}{\emph{Beating the hold-out: bounds for
  k-fold and progressive cross-validation}, }{} in {\em Proceedings of the
  twelfth annual conference on computational learning theory},
  \href{https://doi.org/10.1145/307400.307439}{ COLT '99, (New York, NY, USA),
  203, ACM, 1999}\relax
\mciteBstWouldAddEndPuncttrue
\mciteSetBstMidEndSepPunct{\mcitedefaultmidpunct}
{\mcitedefaultendpunct}{\mcitedefaultseppunct}\relax
\EndOfBibitem
\bibitem{Pivk:2004ty}
M.~Pivk and F.~R. Le~Diberder,
  \ifthenelse{\boolean{articletitles}}{\emph{{sPlot: A statistical tool to
  unfold data distributions}},
  }{}\href{https://doi.org/10.1016/j.nima.2005.08.106}{Nucl.\ Instrum.\ Meth.\
  \textbf{A555} (2005) 356},
  \href{http://arxiv.org/abs/physics/0402083}{{\normalfont\ttfamily
  arXiv:physics/0402083}}\relax
\mciteBstWouldAddEndPuncttrue
\mciteSetBstMidEndSepPunct{\mcitedefaultmidpunct}
{\mcitedefaultendpunct}{\mcitedefaultseppunct}\relax
\EndOfBibitem
\bibitem{LHCb-PAPER-2016-031}
LHCb collaboration, R.~Aaij {\em et~al.},
  \ifthenelse{\boolean{articletitles}}{\emph{{Measurement of the \bquark-quark
  production cross-section in 7 and 13$\tev$ $\proton\proton$ collisions}},
  }{}\href{https://doi.org/10.1103/PhysRevLett.118.052002}{Phys.\ Rev.\ Lett.\
  \textbf{118} (2017) 052002}, Erratum
  \href{https://doi.org/10.1103/PhysRevLett.119.169901}{ibid.\   \textbf{119}
  (2017) 169901}, \href{http://arxiv.org/abs/1612.05140}{{\normalfont\ttfamily
  arXiv:1612.05140}}\relax
\mciteBstWouldAddEndPuncttrue
\mciteSetBstMidEndSepPunct{\mcitedefaultmidpunct}
{\mcitedefaultendpunct}{\mcitedefaultseppunct}\relax
\EndOfBibitem
\bibitem{LHCb-PAPER-2020-018}
LHCb collaboration, R.~Aaij {\em et~al.},
  \ifthenelse{\boolean{articletitles}}{\emph{{Measurement of differential
  $\bquark \bquarkbar$ and $\cquark \cquarkbar$ dijet cross-sections in the
  forward region of $pp$ collisions at $\sqrt{s} = 13$ TeV}},
  }{}\href{https://doi.org/10.1007/JHEP02(2021)023}{JHEP \textbf{02} (2021)
  023}, \href{http://arxiv.org/abs/2010.09437}{{\normalfont\ttfamily
  arXiv:2010.09437}}\relax
\mciteBstWouldAddEndPuncttrue
\mciteSetBstMidEndSepPunct{\mcitedefaultmidpunct}
{\mcitedefaultendpunct}{\mcitedefaultseppunct}\relax
\EndOfBibitem
\bibitem{Skwarnicki:1986xj}
T.~Skwarnicki, {\em {A study of the radiative cascade transitions between the
  Upsilon-prime and Upsilon resonances}}, PhD thesis, Institute of Nuclear
  Physics, Krakow, 1986,
  {\href{http://inspirehep.net/record/230779/}{DESY-F31-86-02}}\relax
\mciteBstWouldAddEndPuncttrue
\mciteSetBstMidEndSepPunct{\mcitedefaultmidpunct}
{\mcitedefaultendpunct}{\mcitedefaultseppunct}\relax
\EndOfBibitem
\bibitem{Wilks:1938dza}
S.~S. Wilks, \ifthenelse{\boolean{articletitles}}{\emph{{The large-sample
  distribution of the likelihood ratio for testing composite hypotheses}},
  }{}\href{https://doi.org/10.1214/aoms/1177732360}{Ann.\ Math.\ Stat.\
  \textbf{9} (1938) 60}\relax
\mciteBstWouldAddEndPuncttrue
\mciteSetBstMidEndSepPunct{\mcitedefaultmidpunct}
{\mcitedefaultendpunct}{\mcitedefaultseppunct}\relax
\EndOfBibitem
\bibitem{CLs}
A.~L. Read, \ifthenelse{\boolean{articletitles}}{\emph{{Presentation of search
  results: The CL$_{\rm s}$ technique}},
  }{}\href{https://doi.org/10.1088/0954-3899/28/10/313}{J.\ Phys.\
  \textbf{G28} (2002) 2693}\relax
\mciteBstWouldAddEndPuncttrue
\mciteSetBstMidEndSepPunct{\mcitedefaultmidpunct}
{\mcitedefaultendpunct}{\mcitedefaultseppunct}\relax
\EndOfBibitem
\bibitem{GammaCombo}
M.~Kenzie {\em et~al.}, \ifthenelse{\boolean{articletitles}}{\emph{{GammaCombo:
  A statistical analysis framework for combining measurements, fitting datasets
  and producing confidence intervals}}, }{}
\newblock
  doi:~\href{https://doi.org/10.5281/zenodo.3371421}{10.5281/zenodo.3371421}\relax
\mciteBstWouldAddEndPuncttrue
\mciteSetBstMidEndSepPunct{\mcitedefaultmidpunct}
{\mcitedefaultendpunct}{\mcitedefaultseppunct}\relax
\EndOfBibitem
\bibitem{LHCb-PAPER-2016-032}
LHCb collaboration, R.~Aaij {\em et~al.},
  \ifthenelse{\boolean{articletitles}}{\emph{{Measurement of the CKM angle
  $\gamma$ from a combination of LHCb results}},
  }{}\href{https://doi.org/10.1007/JHEP12(2016)087}{JHEP \textbf{12} (2016)
  087}, \href{http://arxiv.org/abs/1611.03076}{{\normalfont\ttfamily
  arXiv:1611.03076}}\relax
\mciteBstWouldAddEndPuncttrue
\mciteSetBstMidEndSepPunct{\mcitedefaultmidpunct}
{\mcitedefaultendpunct}{\mcitedefaultseppunct}\relax
\EndOfBibitem
\end{mcitethebibliography}
 
\newpage
\centerline
{\large\bf LHCb collaboration}
\begin
{flushleft}
\small
R.~Aaij$^{32}$,
A.S.W.~Abdelmotteleb$^{56}$,
C.~Abell{\'a}n~Beteta$^{50}$,
F.~Abudin{\'e}n$^{56}$,
T.~Ackernley$^{60}$,
B.~Adeva$^{46}$,
M.~Adinolfi$^{54}$,
H.~Afsharnia$^{9}$,
C.~Agapopoulou$^{13}$,
C.A.~Aidala$^{87}$
% ...

{\footnotesize \it

% ...
$^{9}$Universit{\'e} Clermont Auvergne, CNRS/IN2P3, LPC, Clermont-Ferrand, France\\
% ...
$^{13}$LPNHE, Sorbonne Universit{\'e}, Paris Diderot Sorbonne Paris Cit{\'e}, CNRS/IN2P3, Paris, France\\
% ...
$^{32}$Nikhef National Institute for Subatomic Physics, Amsterdam, Netherlands\\
% ...
$^{46}$Instituto Galego de F{\'\i}sica de Altas Enerx{\'\i}as (IGFAE), Universidade de Santiago de Compostela, Santiago de Compostela, Spain\\
% ...
$^{50}$Physik-Institut, Universit{\"a}t Z{\"u}rich, Z{\"u}rich, Switzerland\\
% ...
$^{54}$H.H. Wills Physics Laboratory, University of Bristol, Bristol, United Kingdom\\
% ...
$^{56}$Department of Physics, University of Warwick, Coventry, United Kingdom\\
% ...
$^{60}$Oliver Lodge Laboratory, University of Liverpool, Liverpool, United Kingdom\\
% ...
$^{68}$Syracuse University, Syracuse, NY, United States\\
% ...
$^{87}$University of Michigan, Ann Arbor, United States, associated to $^{68}$\\
\bigskip
% ...
}
\end{flushleft}

\end{document}